\documentclass[twocolumn]{aastex631}
\usepackage{amsmath}
\usepackage{xspace}

\usepackage{multirow}

\newcommand{\cxo}{{\it Chandra}}
\newcommand{\xmm}{{\it XMM-Newton}\xspace}

\newcommand{\revI}[1]{{{#1}}}

\begin{document}

\title{X-Ray Bright Active Galactic Nuclei in Local Dwarf Galaxies: Insights from eROSITA}

\correspondingauthor{Andrea Sacchi}
\author[0000-0002-7295-5661]{Andrea Sacchi}
\author[0000-0003-0573-7733]{\'Akos Bogd\'an}
\affiliation{Center for Astrophysics $\vert$ Harvard \& Smithsonian, 60 Garden Street, Cambridge, MA 20138, USA}
\author[0000-0003-2521-506X]{Urmila Chadayammuri}
\affiliation{Max Planck Institut f\"ur Astronomie, K\"onigstuhl 17, 69121 Heidelberg, Germany}
\author[0000-0001-5287-0452]{Angelo Ricarte}
\affiliation{Center for Astrophysics $\vert$ Harvard \& Smithsonian, 60 Garden Street, Cambridge, MA 20138, USA}







\begin{abstract}
Although supermassive black holes (SMBHs) reside in the heart of virtually every massive galaxy, it remains debated whether dwarf galaxies commonly host SMBHs. Because low-mass galaxies may retain memory of the assembly history of their black holes (BHs), probing the BH occupation fraction of local dwarf galaxies might offer insights into the growth and seeding mechanisms of the first BHs. In this work, we exploit the Western half of the eROSITA all-sky survey (covering $20,000 \rm{deg^2}$) and compile a catalog of accreting SMBHs in local ($D<200$ Mpc) dwarf galaxies. Cleaning our sample from X-ray background sources, X-ray binaries, and ultraluminous X-ray sources, we identify 74 AGN-dwarf galaxy pairs. Using this large and uniform sample, we derive a luminosity function of dwarf galaxy AGN, fitting it with a power law function and obtaining ${\rm d}N/{\rm d}L_\textup{X} = (15.9\pm2.2)\times L_\textup{X}^{-1.63\pm0.05}$. Measuring the offset between the dwarf galaxies centroids and the X-ray sources, we find that $\approx50\%$ of the AGN are likely off-nuclear, in agreement with theoretical predictions. We compare the BH-to-stellar mass relation of our sample with the local and high-redshift relations, finding that our sources better adhere to the former, suggesting that local AGN across different mass scales underwent a similar growth history. Finally, we compare our sources with semi-analytical models: while our sample's shallowness prevents distinguishing between different seeding models, \revI{we find that the data favor models which keep SMBH in dwarf galaxies active at a moderate rate, motivating model improvement by comparison to AGN in the dwarf galaxy regime}.

\end{abstract}

\keywords{supermasive black holes -- dwarf galaxies -- accretion -- X-ray active galactic nuclei}


\section{Introduction} \label{sec:intro}

Supermassive black holes (SMBHs) are among the most mysterious components of the universe. These SMBHs are found at the center of virtually every massive galaxy and have a profound influence on the evolution of their host galaxies from the earliest epochs to the present day \citep[e.g.][]{croton06,hopkins06a}. Deep optical surveys have revealed a population of optically bright quasars (i.e., SMBHs powered by accretion) at high redshifts ($z \sim 6$), indicating that black holes assembled rapidly, reaching masses of approximately $\sim10^9,{\rm M}_{\odot}$ within the first billion years after the Big Bang \citep[e.g.][]{fan06,willott07,jiang08,mortlock11,venemans13,banados14}. Recently, \textit{Chandra} and \textit{JWST} observations have detected black holes at even higher redshifts, when the universe was only 400-600 million years old \citep{larson23,bogdan24,kokorev23,maiolino24,kovacs24}. Interestingly, these SMBHs are located in low-mass galaxies that exhibit similarities to present-day dwarf galaxies.

\begin{figure*}[ht!]
\centering
\includegraphics[width = 0.8\textwidth]{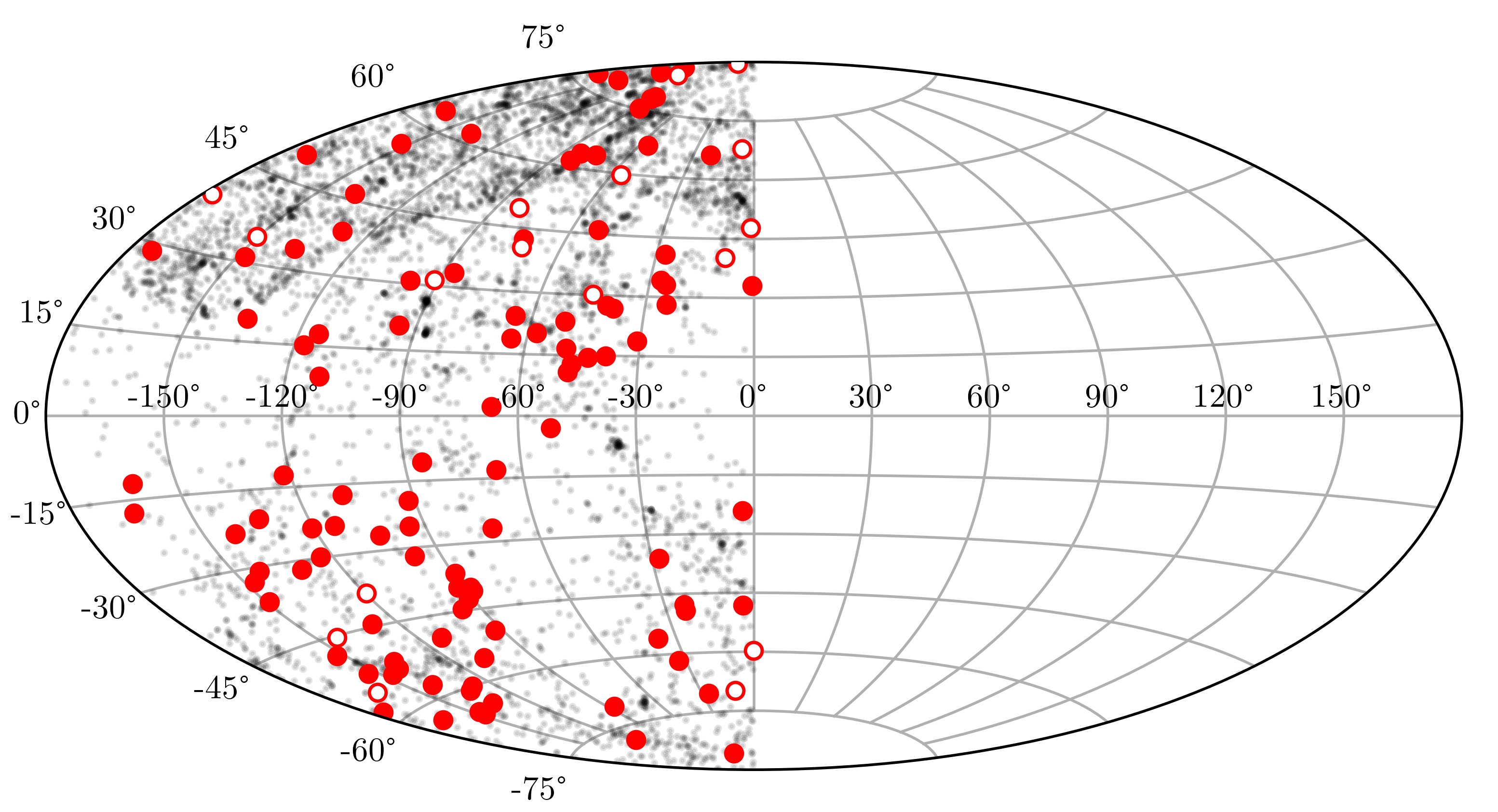}
\caption{Aitoff projection of the source in the parent HECATE sample (in grey) and the matches with eRASS1 (red markers). The sources in our final sample are represented by filled red dots while the sources identified in the MILLIQUAS catalog and excluded from our sample are indicated by empty red circles. \label{fig:location}}
\end{figure*}

The formation of SMBHs is a particularly intriguing question, with several promising pathways suggested: remnants of Population III stars \citep{bromm03}; mergers in nuclear star clusters \citep{portegies04,gurkan04,natarajan21}; direct collapse of pre-galactic gas disks where fragmentation and star formation are suppressed \citep{lodato06,begelman06}; or dark stars powered by dark matter annihilation \citep{spolyar08}. There are two promising avenues to distinguish between various theoretical models. One approach is to observe SMBHs in the early universe ($z\gtrsim9$), close to their seeding epoch. Alternatively, we can investigate the occupation fraction and properties of SMBHs in dwarf galaxies at the present epoch.

Studying the population of SMBHs in nearby dwarf galaxies offers a powerful and complementary method to probe the seeding mechanisms and growth history of the first black holes. While SMBHs in today’s massive galaxies ($>10^{10} \ \rm{M_{\odot}}$) did not retain a memory of their origins due to the intense merger activity of their host galaxies, the black hole occupation fraction of local dwarf galaxies ($10^{8.5-9.5} \ \rm{M_{\odot}}$) is sensitive to the black hole seeding mechanism \citep{volonteri10}. If SMBHs formed through the collapse of Population III stars, every dwarf galaxy in the local universe is expected to host an SMBH. However, if black holes formed through the direct collapse of massive gas clouds, only about half of present-day dwarf galaxies are predicted to harbor an SMBH (see e.g. \citealt{greene12}).

Over the past years, various methods have been employed to detect accreting SMBHs (or Active Galactic Nuclei -- AGN) in dwarf galaxies (see \citealt{mezcua17,greene20,reines22} and references therein). The most common approach for systematically identifying AGN in dwarf galaxies is based on optical spectroscopy, utilizing emission-line diagnostics to distinguish between star formation-related and accretion-powered (i.e., AGN) activity \citep{greene04,greene07,reines13,moran14,baldassare15,chilingarian18}. While highly efficient, identifying $\gtrsim600$ AGN in dwarf galaxies to date, this technique is biased towards objects accreting at a high Eddington ratio and those hosted in solar and super-solar metallicity galaxies (e.g., \citealt{greene20}; but see \citealt{polimera22} and \citealt{mezcua24} for ways to correct for these effects).

Alternative methods to identify SMBHs in dwarf galaxies rely on optical, UV, and infrared variability \citep{baldassare18,baldassare20,martinez20,ward22,arcodia24}, coronal lines \citep{satyapal07,cann18,cann21,salehirad22,reefe23}, radio emission \citep{mezcua19,reines20,davis22}, and X-ray observations \citep{schramm13,lemons15,pardo16,mezcua18,birchall20,latimer21,ohlson24,bykov24}. These last two methods are particularly effective in identifying AGN due to the high nuclear-to-host contrast that these bands offer: high angular resolution radio observations allow the separation of the AGN from other radio-emitting components of the galaxy, and the X-ray spectrum of AGN is distinct from other stellar X-ray sources \citep{ho08,merloni16}. However, a sensitive volume-limited radio survey of dwarf galaxies in the nearby universe is prohibitively expensive \citep{reines20}. Similarly, X-ray observational campaigns have, to date, been either too shallow and with poor angular resolution, or too limited in sky coverage.

The game-changer in this domain is eROSITA (extended ROentgen Survey with an Imaging Telescope Array, \citealt{predehl21}), the main instrument aboard the German-Russian satellite \textit{Spektrum Roentgen Gamma} (\textit{SRG}) \citep{sunyaev21}. eROSITA provides today's most sensitive all-sky survey in the X-ray band, and eRASS1, the first data release, comprises more than 930,000 individual sources across the Western half of the sky \citep{merloni24}. In this work, we leverage the opportunity offered by eRASS1 to build a sample of uniformly selected AGN in local dwarf galaxies, thereby probing their \revI{properties.}

\revI{Similar efforts have} been conducted by \citet{bykov24}, who utilized eROSITA data from the Eastern half of the sky and the MPA-JHU galaxy catalog, based on the SDSS DR12 \citep{alam15}, \revI{and \citet{hoyer24}, who focused on massive BHs hosted in nuclear star clusters.} 
\revI{Based on the same parent samples we adopted, i.e. eRASS1 and the HECATE catalog \citep{kovlakas21}, but focused on an entirely different science goal, is the effort performed by \citet{kyritsis24}. They studied the star-formation-driven X-ray emission of normal galaxies, which we instead excluded from our sample which collects only accretion-powered sources.} 


The paper is organized as follows: in Section \ref{sec:sample} we describe the sample-building procedure and cleaning criteria; in Section \ref{sec:results} we analyze the properties of the sources in our sample; in Section \ref{sec:mod} we compare our sample with semi-analytical predictions and in Section \ref{sec:conc} we draw our conclusions. 

\section{Sample selection and properties} 
\label{sec:sample}
\subsection{Defining the galaxy sample}

To search for AGN candidates located in dwarf galaxies within the eRASS1 footprint, we cross-matched the positions of approximately 930,000 X-ray sources in the eRASS1 catalog \citep{merloni24} with those of local dwarf galaxies. To define the sample of nearby dwarf galaxies, we relied on the HECATE catalog \citep{kovlakas21}, an all-sky, value-added catalog that includes information about the size, distance, and star-formation rate of more than 200,000 local galaxies. It is more than $50\%$ complete in terms of B-magnitude up to $\approx170$~Mpc. Filtering the HECATE catalog for local dwarf galaxies (within $D<200$ Mpc and with stellar masses of $10^6~{\rm M_\odot}<M_{\rm gal}<3\times10^9~{\rm M_\odot}$) resulted in a sample of 5,775 objects. We adopted a radius of $15\arcsec$ to cross-match the positions of eRASS-detected X-ray sources with the dwarf galaxies. This radius corresponds to $\approx10.5$~kpc at the mean distance of the galaxy sample, which is approximately three times the mean radius of the dwarf galaxies. We thus obtained 120 X-ray source-dwarf galaxy pairs. The location of the sources in the parent HECATE sample and the results of our cross-match are shown in Figure~\ref{fig:location}.

\subsection{Chance alignment with background AGN}

\begin{figure}[t!]
\centering
\includegraphics[width = 0.5\textwidth]{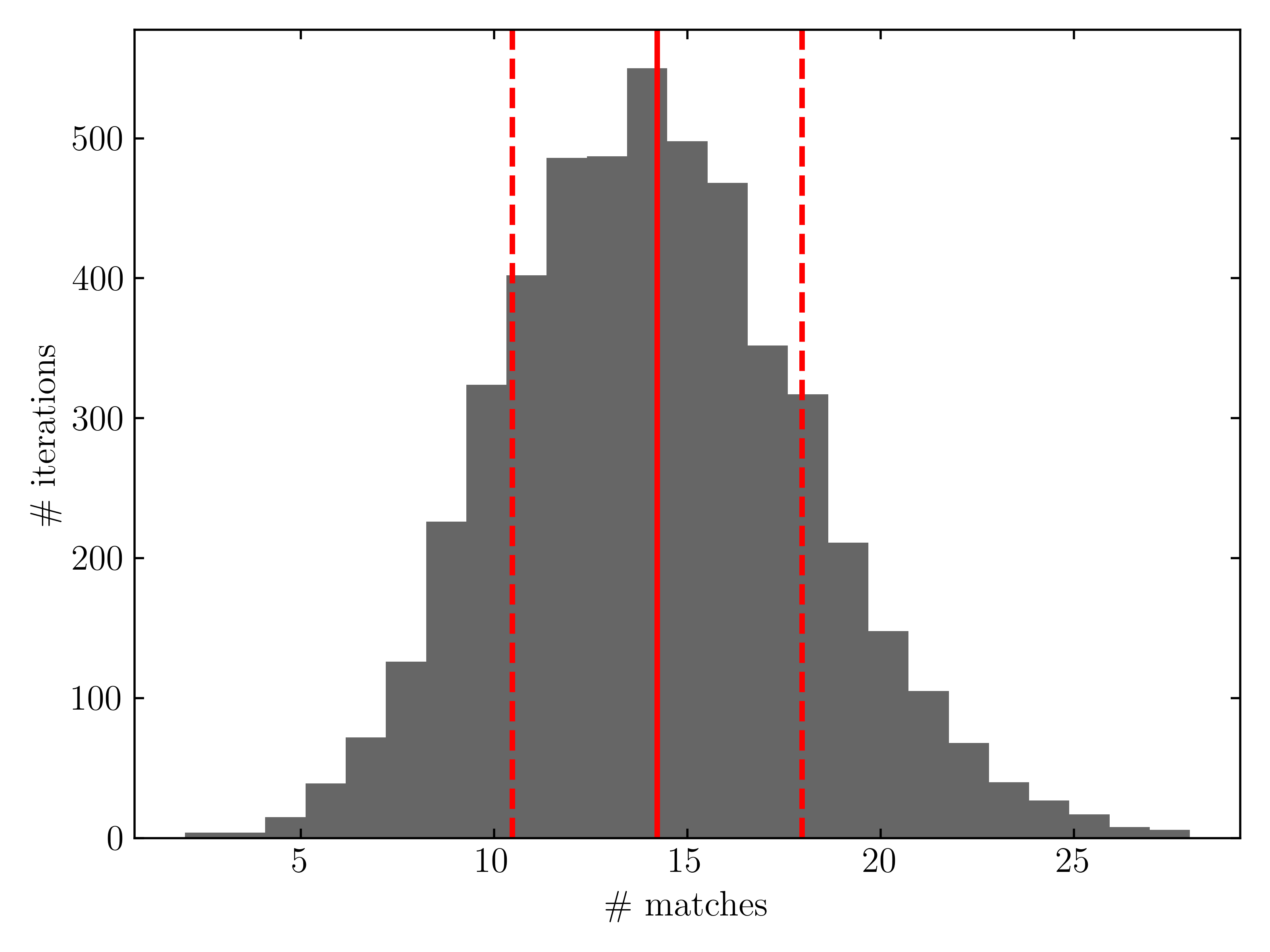}
\caption{Histogram of the number of matches obtained in 5,000 iterations of randomly generated coordinated and the eRASS1 catalog, the vertical red lines indicate the average number of matches (solid) and the $1\sigma$ uncertainty interval (dashed).
\label{fig:mc}}
\end{figure}

A fraction of these matches likely result from chance alignments between dwarf galaxies and background AGN: cosmic X-ray background sources could be projected at the positions of the dwarf galaxies. We used two approaches to determine the number of such sources. First, we carried out 5,000 Monte Carlo simulations by generating 5,775 random coordinates representing the dwarf galaxy population in these simulations. Then we cross-matched the generated coordinates with the eRASS1 catalogue adopting the same match radius of $15\arcsec$. On average, we found $14\pm4$ matches which can be attributed to chance alignment. The distribution of the matches obtained through the Monte Carlo procedure is shown in Figure~\ref{fig:mc}. 

Second, we compared the locations of the X-ray source-galaxy pairs to the MILLIQUAS catalog \citep{flesch23}, which includes nearly one million quasars. This comparison yielded 17 matches, aligning well with the results of the Monte Carlo simulations. We then removed these 17 sources from our sample, reducing it to 103 X-ray source-dwarf galaxy pairs.

\subsection{Unresolved X-ray binaries}

The population of X-ray binaries can significantly contribute to the X-ray emission of galaxies. Due to the broad PSF of eROSITA and the shallow depth of the eRASS1 survey, distinguishing these X-ray binaries from AGN emission is challenging. Because the collective emission from X-ray binaries correlates with the galaxy's stellar mass (for low-mass X-ray binaries) and star-formation rate (for high-mass X-ray binaries), we probed whether the population of X-ray binaries could account for the X-ray emission observed in the studied dwarf galaxies. To this end, we used the galaxy-wide X-ray binary emission scaling relation \citep{lehmer10}:  
$$L_{2-10\,{\rm keV}}\,{\rm (erg/s)}=\alpha M_\textup{gal}+\beta{\rm SFR} \ ,$$ 
where $\alpha=(9.05\pm0.37)\times10^{28}\,{\rm erg/s/M_\odot}$ and $\beta=(1.62\pm0.22)\times10^{39}\,{\rm erg/s/(M_\odot/y)}$. This relation has a dispersion of 0.34~dex.

The HECATE catalog provides stellar masses for all galaxies in our sample and SFR values for 47 of them. For the remaining sources, we computed their SFR using the relation suggested by \citet{rieke09}:
${\rm SFR}=7.8\times10^{-10}\,L_{24\,\mu}/\rm{L}_\odot {\rm (erg/s)}$. 
For 57 sources, we used the $22,\mu$ luminosity (W4 {\em Allwise} band) instead of the $24,\mu$ luminosity, and for two galaxies, we used the $25,\mu$ luminosity (IRAS 25), assuming that $L_{24,\mu}/L_{25,\mu}\approx L_{24,\mu}/L_{22,\mu}\approx1$. Finally, for two galaxies without infrared photometry, we assumed the average SFR of our parent sample.

The formula provided by \citet{lehmer10} derives the luminosity in the $2-10$~keV band, so we converted the $0.2-2.3$ keV luminosity provided by the eRASS1 catalog to the $2-10$ keV band assuming a Galactic absorption of $3\times10^{20}$~cm$^{-2}$ and a power law with photon index \revI{$\Gamma=2$}.

For 15 galaxies, the X-ray emission from the eRASS1 sources is comparable to the X-ray emission expected from the population of unresolved X-ray binaries. For the remaining galaxies, the X-ray sources are more than 2.2 times brighter than expected from the population of X-ray binaries, making it unlikely that the eRASS-detected X-ray sources originate from X-ray binaries.

\begin{figure}[t!]
\centering
\includegraphics[width = 0.5\textwidth]{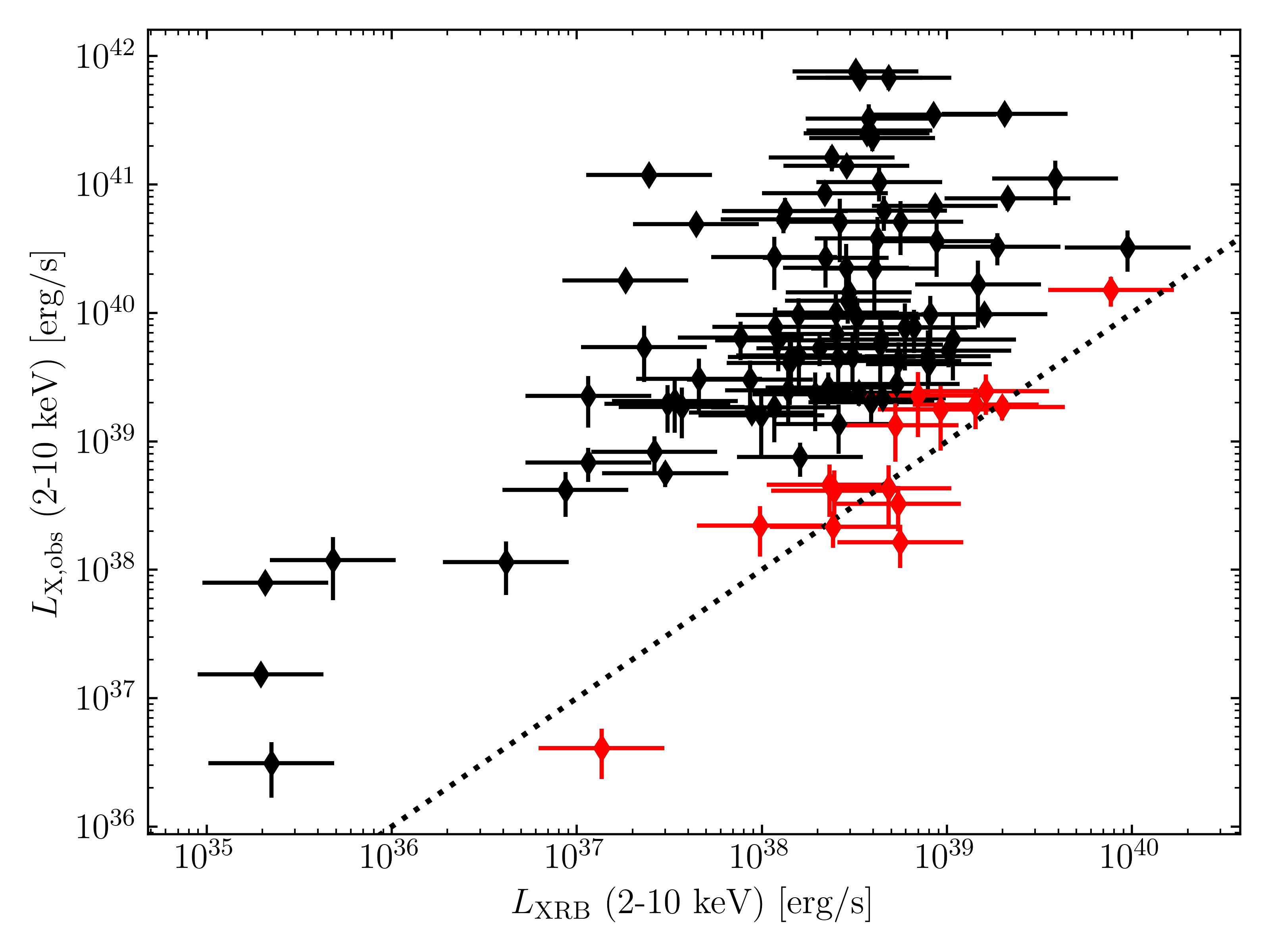}
\caption{Observed vs expected X-ray luminosity from X-ray binaries for our sample of dwarf galaxies. The dashed line indicates the bisectrix. The cumulative emission from X-ray binaries may account for the observed eROSITA X-ray emission in the sources highlighted in red. These galaxies were removed from our sample.
\label{fig:sfr}}
\end{figure}

\subsection{Resolved ultra-luminous X-ray sources}

The sources in the high-luminosity tail of the distribution of low-mass and high-mass X-ray binaries (LMXBs/HMXBs), appearing as single, resolved sources, might still contaminate our sample of X-ray source-dwarf galaxy pairs. These X-ray sources are expected to dominate the low-luminosity tail of our sample. To be conservative, we removed all sources from our sample with luminosity $L_{\rm X}<10^{39} \ \rm{erg \ s^{-1}}$ in our reference $0.2-2.3$~keV band. Assuming all these were HMXBs, we then inferred the number of residual contaminants in the rest of the sample.

\begin{figure}[t!]
\centering
\includegraphics[width = 0.5\textwidth]{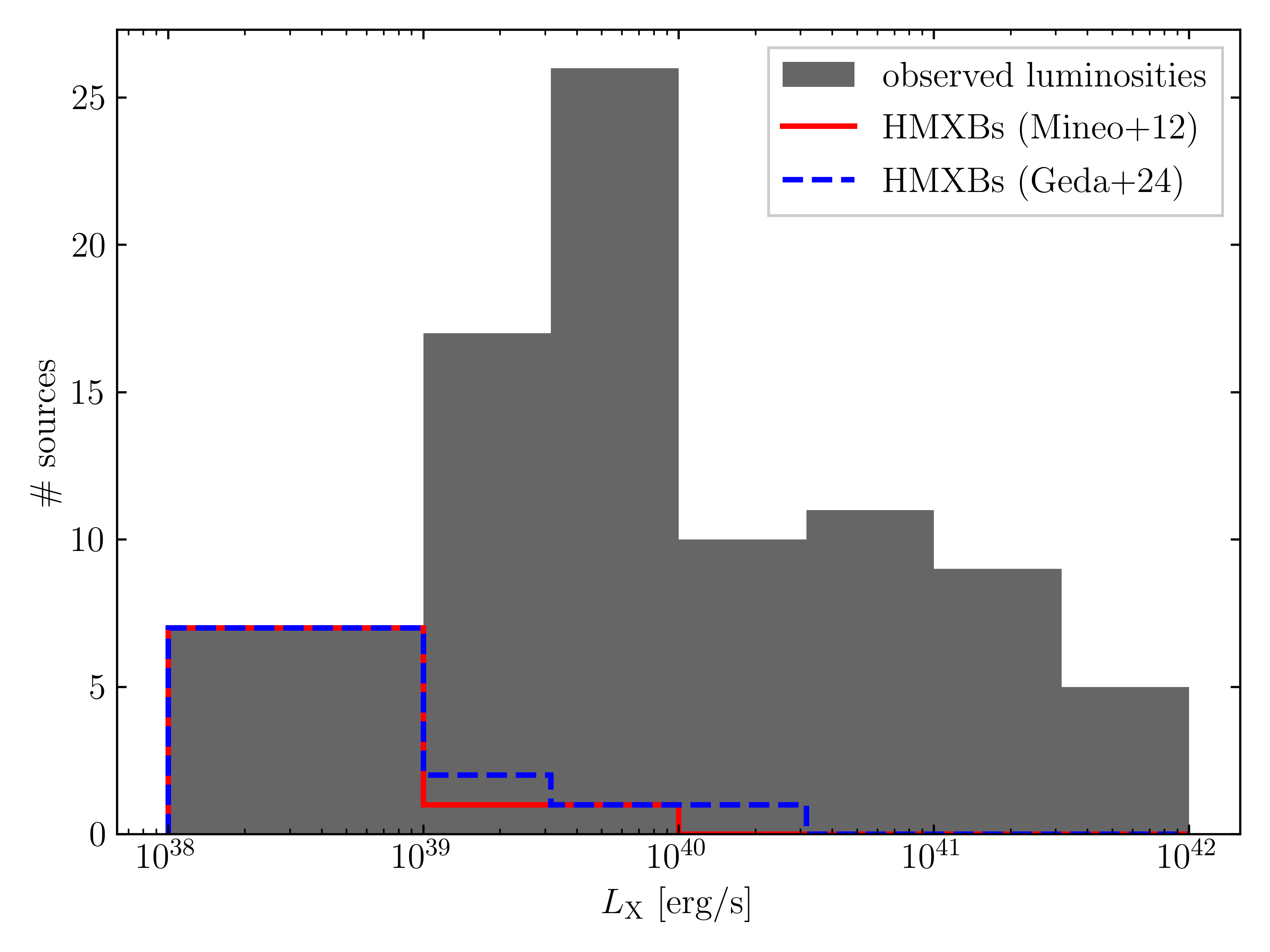}
\caption{Distribution of the observed X-ray luminosity of the AGN in our sample. Dashed and dotted lines indicate the predicted HMXBs populations normalized to the number of sources with a luminosity $L_{\rm X}<10^{39} \ \rm{erg \ s^{-1}}$.
\label{fig:lum_poll}}
\end{figure}

To estimate this number, we adopted the HMXB luminosity functions by \citet{mineo12a} and \citet{geda24}. The former better samples the high-luminosity end of the distribution, as it is based on a larger sample of sources, but it is not calibrated for dwarf galaxies. Conversely, the latter is calibrated for dwarf galaxies but is heavily affected by stochasticity at its high-luminosity end. Figure \ref{fig:lum_poll} shows the luminosity distribution of the X-ray sources in our sample, with the high-mass X-ray binary luminosity functions superimposed, normalized for the number of sources with $L_{\rm X}<10^{39} \ \rm{erg \ s^{-1}}$. Based on these models, we expect that our final sample of 78 sources will be contaminated by $2–4$ HMXBs with a luminosity exceeding $10^{39} \ \rm{erg \ s^{-1}}$. Even accounting for an incompleteness fraction of about $50\%$ (see Section \ref{sec:xlf} for further details), the number of expected contaminants is $\lesssim6$.

\begin{figure*}[!t]
\centering
\includegraphics[width=0.32\textwidth]{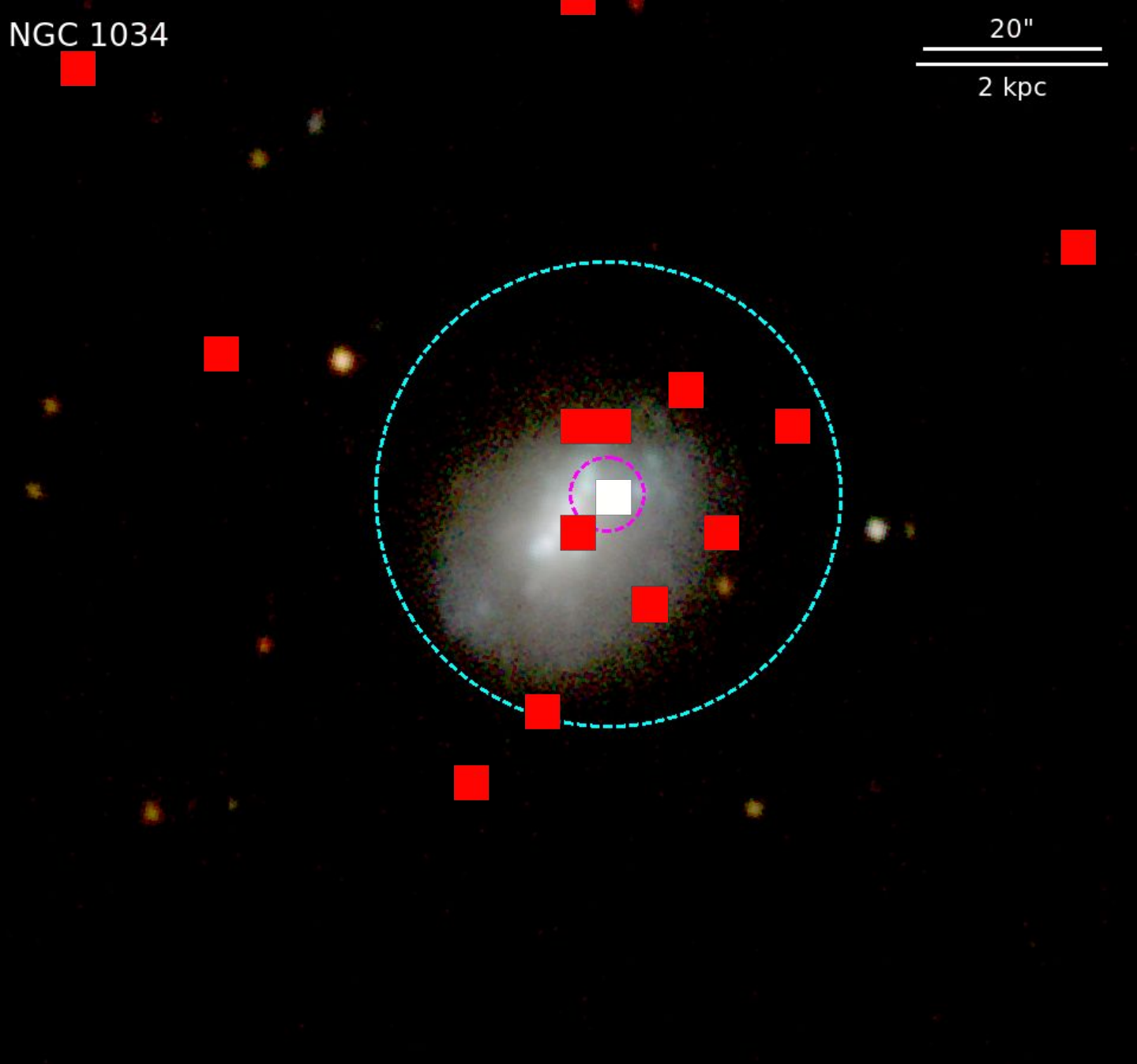}
\includegraphics[width=0.32\textwidth]{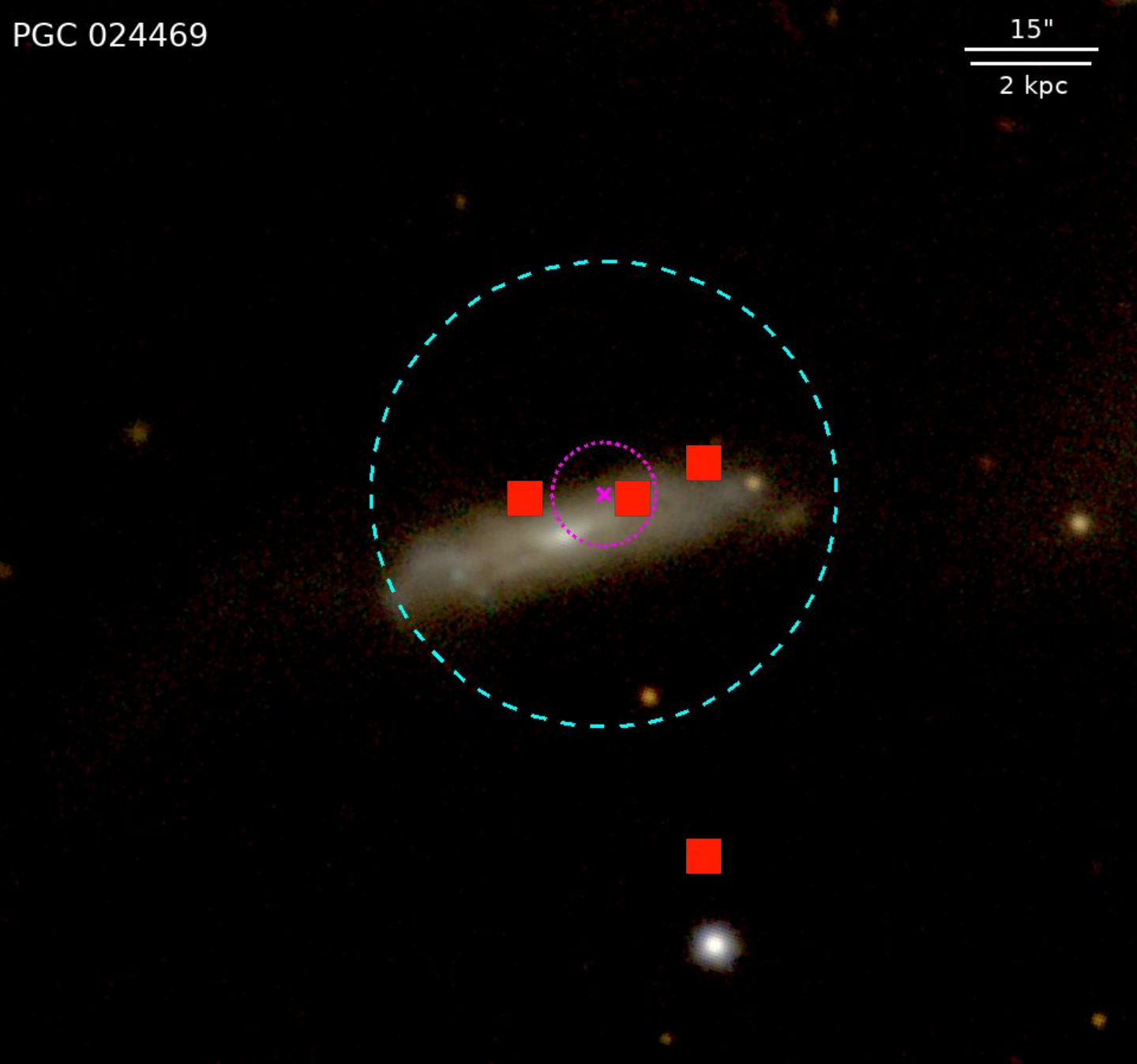}
\includegraphics[width=0.32\textwidth]{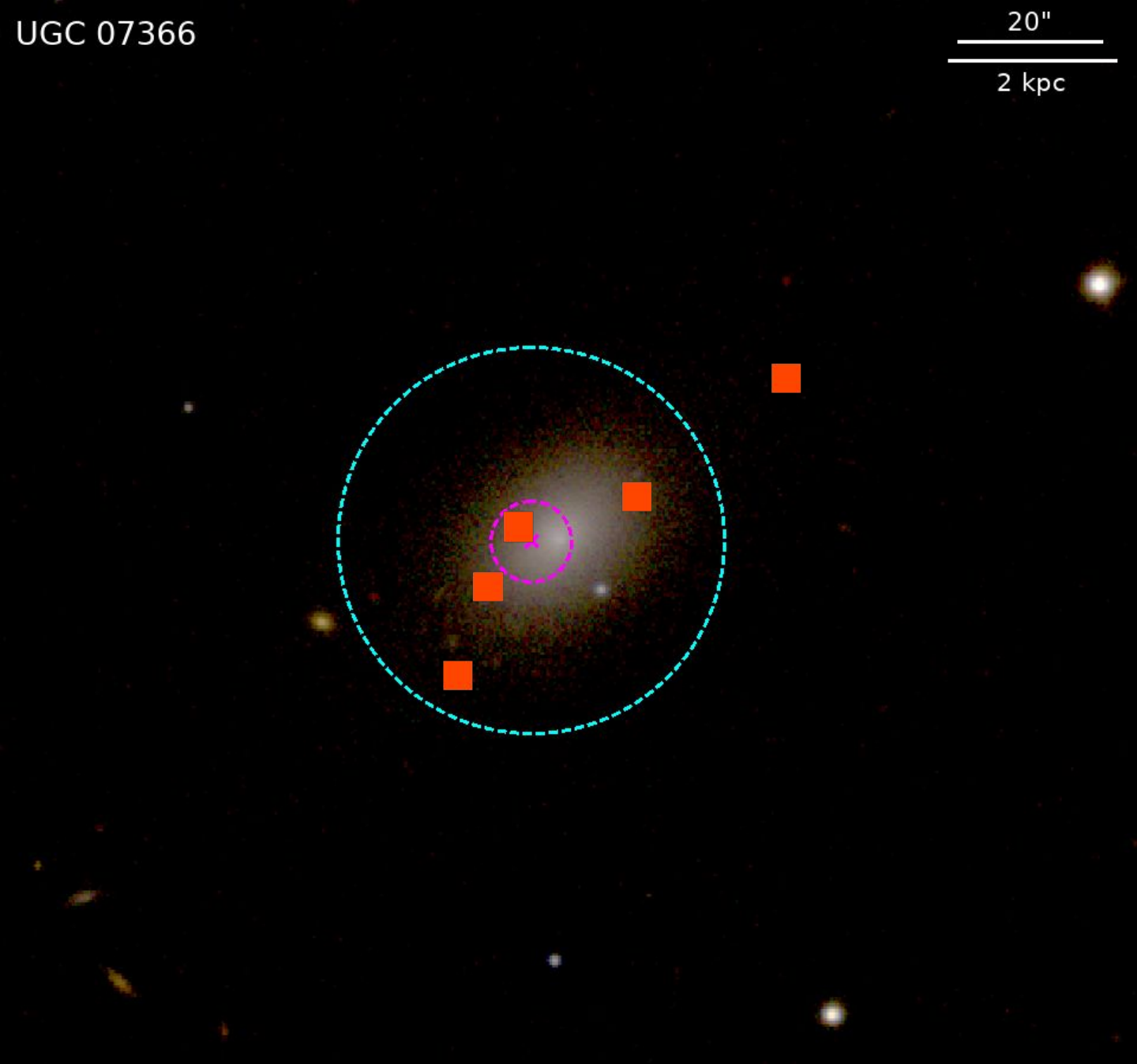}
\caption{A representative subset of our sample of AGN hosted in local dwarf galaxies. The three-color composite images were obtained using the \textit{SDSS} \textit{gri}-filters. The X-ray counts are shown with the white (two counts) and red (one count) pixels. The cross and the small magenta circle indicate the centroid and positional uncertainty of the X-ray source taken from the eRASS1 point source catalog \citep{merloni24}, respectively \revI{(typically around $5\arcsec$)}. The large cyan circle shows the PSF size ($r=26\arcsec$) of eROSITA in scanning mode.}
\label{fig:ero}
\end{figure*}

\subsection{Crossmatch with archival X-ray data}

A subset of the sources in our sample was previously observed with \textit{XMM-Newton} and/or \textit{Chandra}, and data are available in the respective archives\footnote{\url{https://nxsa.esac.esa.int/nxsa-web/\#search}, \url{https://cda.harvard.edu/chaser/mainEntry}}. One source represents a spurious detection, not associated with a point-like source but rather with the diffuse emission of a nearby galaxy cluster, and was hence removed from our sample. Four sources in our sample are identified as ultraluminous X-ray sources (ULXs), i.e.\ HMXBs with a luminosity exceeding $10^{39} \ \rm{erg \ s^{-1}}$, based on their multi-wavelength emission. 

Three sources were identified as candidate AGN due to their infrared and X-ray emission, and their nature has been further investigated with \cxo\ pointed observations. In two cases, the \cxo\ observations revealed compact sources of hard X-ray photons in the nuclear regions of the host galaxies, confirming the AGN nature of their emission \citep{dudik05,latimer21}. In one case, however, the compact source lies outside the nuclear region of its host galaxy. While this might suggest that the source is a ULX \citep{thygensen23}, it is also possible that it is an off-nuclear AGN, as a fraction of dwarf galaxies ($\sim50\%$) are expected to host wandering SMBHs outside the nuclear region \citep{reines20,bellovary21}.

Finally, five additional sources fall within the footprints of either \textit{XMM-Newton} or \textit{Chandra} observations. While detected as point-like sources, they are too far off-axis for their locations to be precisely assessed within their host galaxies, although their X-ray spectra are consistent with being AGN. A detailed description of all the sources mentioned above is presented in Appendix~\ref{app:xray}.

\subsection{Final sample of dwarf galaxy-X-ray source pairs}

\begin{figure}[t!]
\centering
\includegraphics[width = 0.5\textwidth]{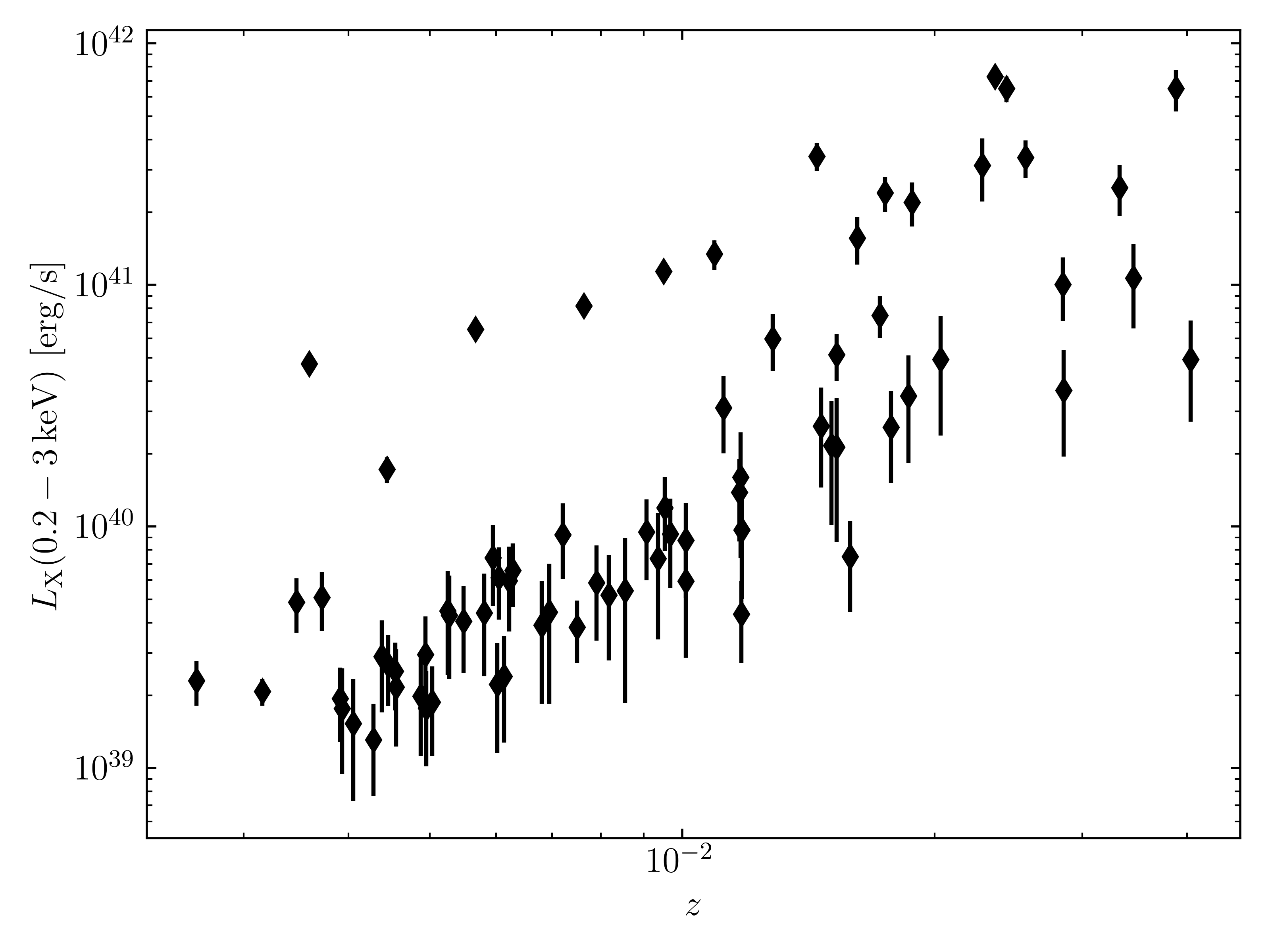}
\caption{X-ray luminosity in the $0.2-2.3$~keV eRASS1 band as a function of the redshift of each source in our final sample of 74 AGN.
\label{fig:dlum}}
\end{figure}

Our final sample, after removing the four known ULXs, is composed of 74 dwarf galaxies with an eROSITA-detected source, \revI{which corresponds to a sample-average AGN fraction of $74/5775\approx1.3\%$.} In Figure~\ref{fig:dlum} we show the X-ray luminosity of the sources as a function of the redshift of the galaxies. Postcard images of a representative set of three of these candidates are shown in Figure~\ref{fig:ero}. All the relevant properties of each dwarf galaxy-X-ray source pair are reported in Table~\ref{tab:prop}.

\section{Results}
\label{sec:results}

\subsection{X-ray hardness ratios}

To further characterize the properties of the eROSITA-detected X-ray sources, we first probe their spectral properties. Because the bulk of the sources in our sample only have a handful of X-ray counts, we cannot obtain reliable hardness ratios for most of the individual sources. Therefore, we compute the sample-average X-ray hardness ratio. To this end, we define the hardness ratio as $\rm{HR}=S/M$ where the number of counts is measured in the soft (S=$0.5-1$ keV) and medium bands (M=$1-2$ keV), respectively. We obtained a hardness ratio of S/M=$0.98_{-0.06}^{+0.04}$, where the uncertainties were computed using a Bayesian approach with the \texttt{BEHR} code \citep{BEHR06}. Considering the eROSITA effective area curve, this measured value is compatible with a power-law model with a slope of $\Gamma=1.8$ seen through a column density of $N_\textup{H}=3\times10^{20}$~cm$^{-2}$. This spectral model is in good agreement with the typical slope of AGN detected in the eROSITA early data release (eFEDS, \citealt{liu22}) and disfavor the scenario in which the X-rays we detect come from diffuse emission.

\subsection{Off-nuclear SMBHs}

Another key piece of information that helps us better characterize our sample and compare our results with theoretical predictions is the offset between the X-ray source and the host galaxy centroid. As argued for one of the sources with a \cxo\ follow-up observation, the most natural explanation for an off-nuclear X-ray bright source is a ULX scenario (see e.g. \citealt{kaaret17,fabrika21} for reviews on the topic). However, not only do we detect sources in excess of the ones expected from HMXB/ULX luminosity functions, but given the shallow gravitational potential of dwarf galaxies, the fraction of off-nuclear AGN is expected to be as large as one-third \citep{reines20,bellovary21}. 

\begin{figure}[t!]
\centering
\includegraphics[width = 0.5\textwidth]{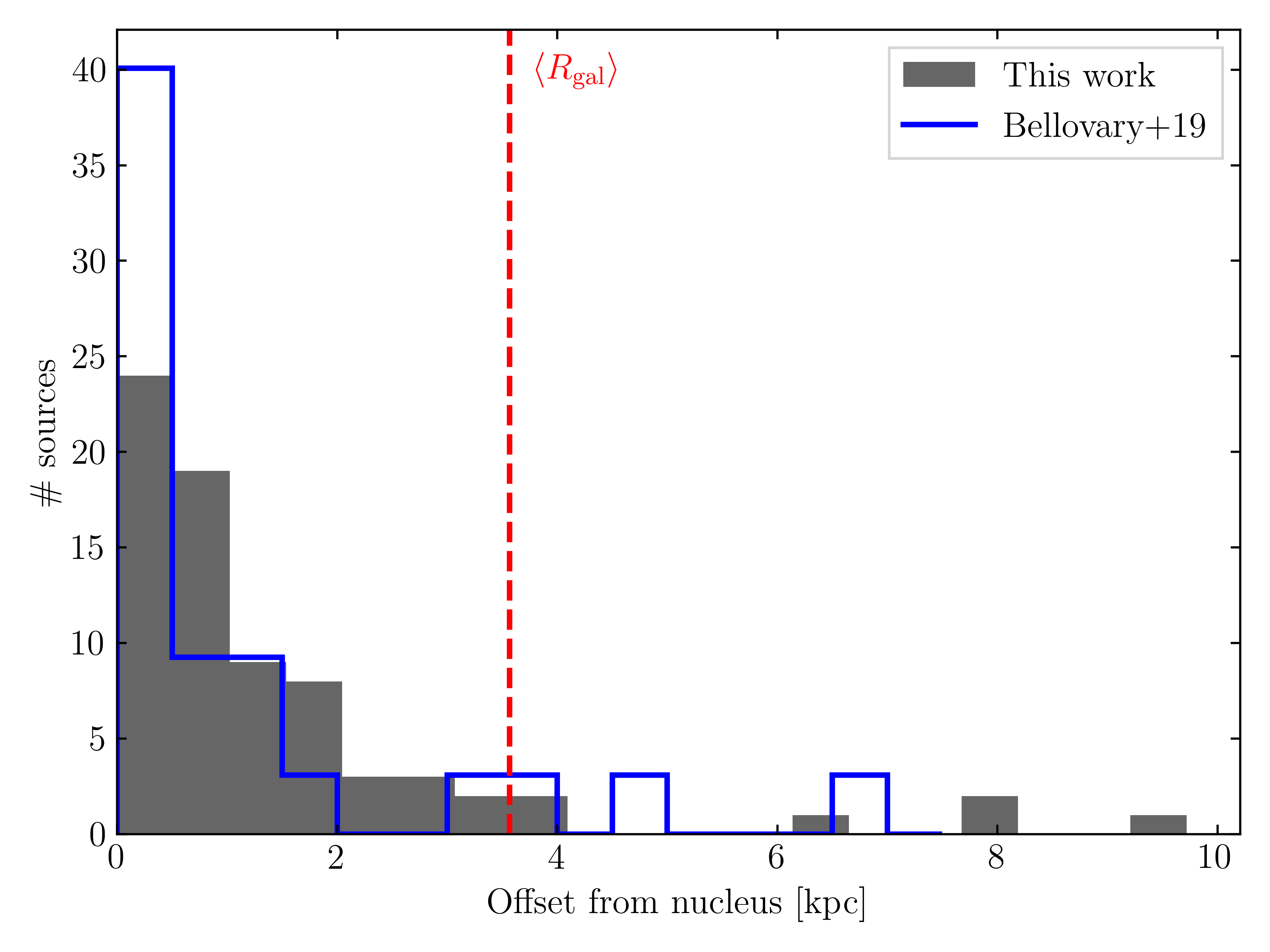}
\caption{Offset of the X-ray source position with respect to the host galaxy centroid in kpc. The red dashed line indicates the average radius of the dwarf galaxies in our sample. The blue histogram shows the prediction by \citet{bellovary19}, rescaled to match our sample size.
\label{fig:offset}}
\end{figure}

Figure \ref{fig:offset} shows the offset of the X-ray source position with respect to its host galaxy centroid. The bulk of our sample has a small offset, with only about 10 sources at a distance larger than $2$~kpc from their host galaxy centroid. Once the positional error of the X-ray sources (\texttt{POS\_ERR} from the eRASS1 catalog) is taken into account, the location of the X-ray source is compatible with the nucleus of its host galaxy for about $50\%$ of the objects in our sample, and for more than $80\%$, it is compatible within two times the positional error.

These results align with theoretical predictions and observational evidence \citep{bellovary19,reines20,bellovary21} for AGN in dwarf galaxies (as also shown in blue in Figure~\ref{fig:offset}), although some caveats must be considered. The large PSF of eROSITA in scanning mode ($\approx26\arcsec$) prevents \revI{any conclusive statement on possible source blending}, and dwarf galaxies often exhibit irregular morphology, making it sometimes difficult to identify their centers \citep{pasetto03,lazar24}. To conclusively determine the \revI{nature and location} of the X-ray sources, follow-up high spatial resolution observations with \textit{Chandra} would be needed. 

\subsection{X-ray luminosity function}\label{sec:xlf}

To further characterize the sources, infer properties of the bulk population of AGN in dwarf galaxies, and compare them with populations of other X-ray emitting sources, we constructed the X-ray luminosity function of our sample. To correct for the eROSITA incompleteness, we weighted each luminosity bin by the probability of detecting an X-ray source in that bin. The probability of detecting a source with luminosity $L_\textup{X}$ was computed by simulating $N$ sources distributed in redshift like the sources in our parent sample. Using the provided sensitivity map, we calculated how many sources, $D$, eROSITA would detect. The probability of detecting a source with luminosity $L_\textup{X}$ is simply $D/N$.

\begin{figure}[t!]
\centering
\includegraphics[width = 0.5\textwidth]{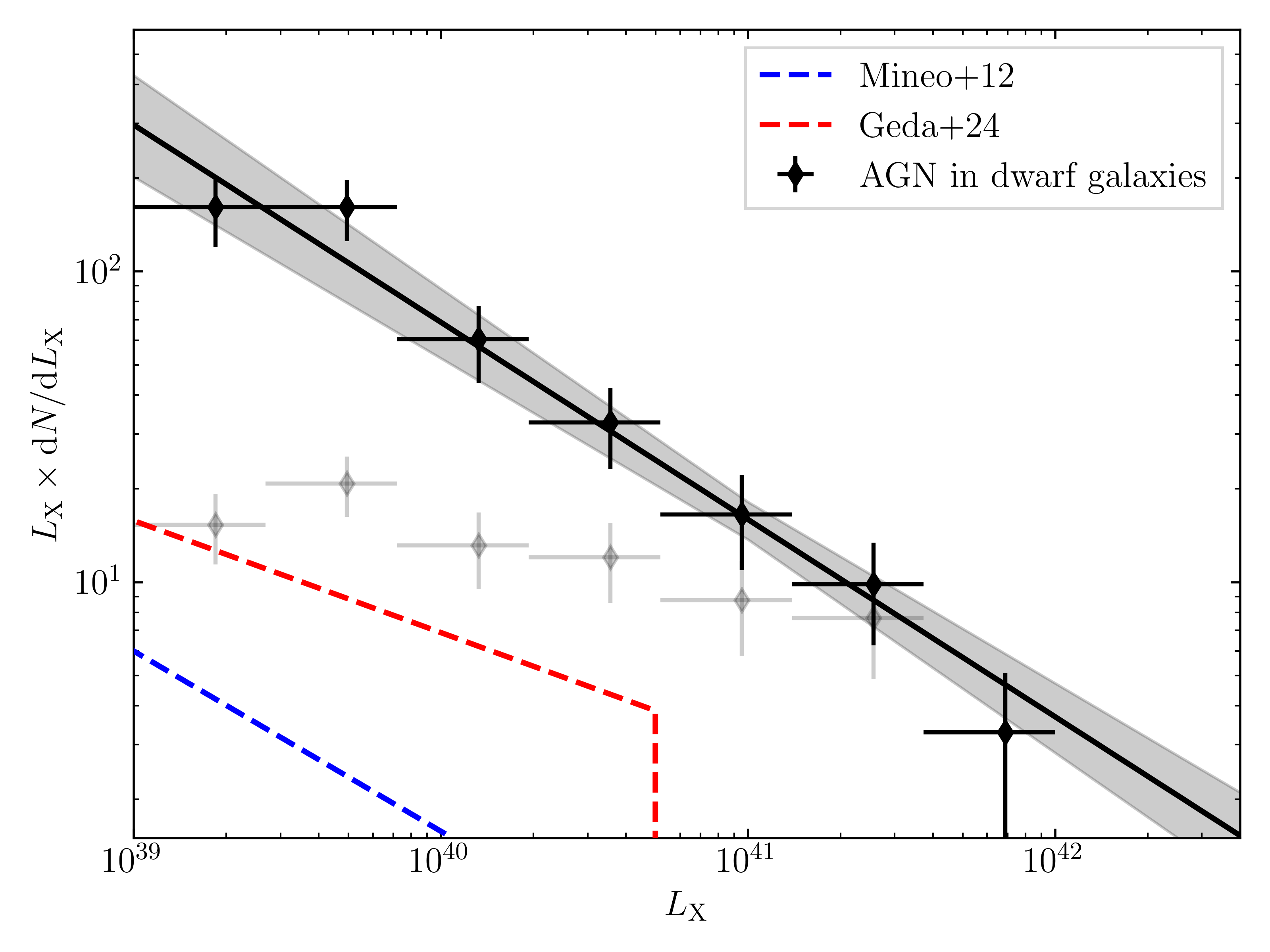}
\caption{X-ray luminosity function of our sample of AGN in dwarf galaxies (black data points), the raw luminosity function, before the completeness correction is also shown (grey data points). The black solid line and grey shaded region indicate our best fit and its dispersion. The blue and red dashed lines indicate the luminosity functions of HMXBs/ULXs \citep{mineo12a,geda24}.
\label{fig:xlf}}
\end{figure}

The X-ray luminosity function we obtained is shown in Figure~\ref{fig:xlf}. We also show the luminosity functions of HMXBs/ULXs from previous works \citep{mineo12a,geda24}. A direct comparison between these and the luminosity function derived from our sample clearly shows that X-ray source-dwarf galaxy pairs in our sample represent a population of sources with significantly different behavior in terms of X-ray emission. We fit our luminosity function with a power-law model and obtained the following:
\begin{equation}
    {\rm d}N/{\rm d}L_\textup{X} = (15.9\pm2.2)\times L_\textup{X}^{-1.63\pm0.05},
\end{equation}
where $L_\textup{X}$ is expressed in units of $10^{41} \ \rm{erg \ s^{-1}}$. The luminosity function we obtained, although slightly steeper, is compatible with those reported by \citet{birchall20} (based on \xmm\ X-ray data) and \citet{bykov24} (based on the eROSITA data of the Eastern sky), both of which used SDSS data as dwarf galaxies parent sample. The difference might stem from the slightly different selection criteria and the luminosity range we considered: both authors report a flattening of the luminosity function below $10^{39} \ \rm{erg \ s^{-1}}$, which we are not sensitive to as we excluded all sources below that threshold. 

\subsection{Galaxy-to-black hole mass relation}

To place our results in a broader context, we compare the black hole mass ($M_\bullet$) of the AGN in our sample, inferred from their X-ray luminosity, to the mass of their host galaxies ($M_\star$). The recent detection of AGN at high redshift, powered by SMBHs with a mass equal to $\approx10-100\%$ of their host galaxy \citep{pacucci23,bogdan24,maiolino24}, has prompted theoretical efforts to reconcile these observations with the local $M_\bullet-M_\star$ relation. This local relation typically accounts for SMBHs with a mass $\approx0.1\%$ of the galaxies they inhabit \citep{reines15}. It might be possible to reconcile this discrepancy by invoking observational biases, such as the difficulty in detecting high-redshift, low-mass (and hence low-luminosity) AGN \citep{li24b}.

\begin{figure*}[t!]
\centering
\includegraphics[width = 0.85\textwidth]{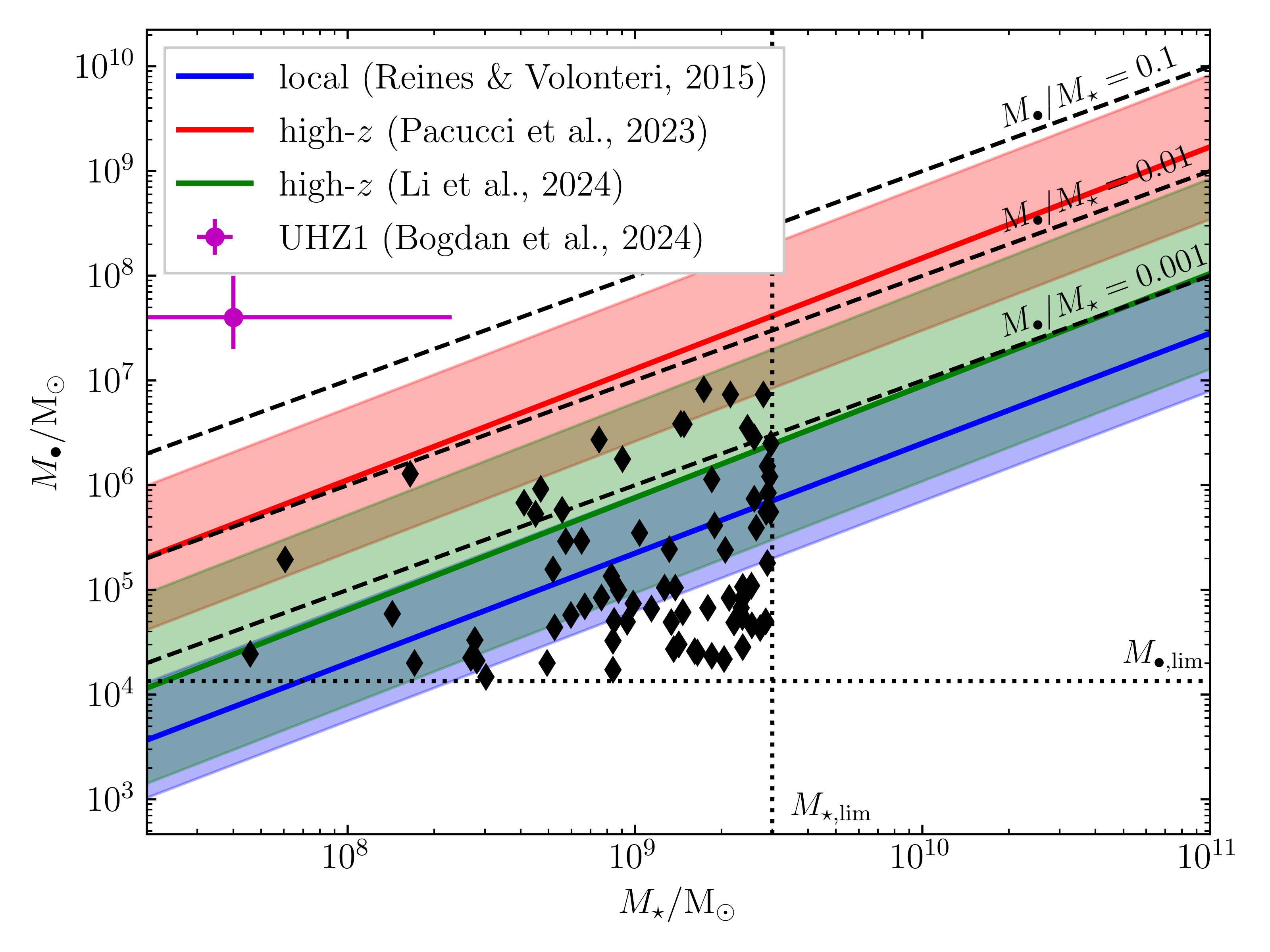}
\caption{$M_\bullet-M_\star$ for the AGN in our sample. The blue and red lines indicate the low- and high-redshift relations, respectively, with the shaded regions indicating their dispersion. The dotted lines indicate the limit of the parameter space we explored in our analysis: the lower limit on the detectable BH mass, corresponding to a luminosity limit of eRASS1 at the average distance of our sample and the upper limit of the galaxy mass. The magenta marker indicates UHZ1, a $z=10.1$ galaxy with a \textit{Chandra}-detected accreting over-massive black hole \citep{bogdan24}. 
\label{fig:mhms}}
\end{figure*}

Exploiting the analogies between high-redshift AGN and AGN hosted in low-mass galaxies, we further investigate these effects by studying the $M_\bullet-M_\star$ relation in our dwarf galaxy sample. We infer the SMBH mass from the X-ray luminosity of the AGN. To this end, we employ the bolometric correction $L_\textup{bol}/L_\textup{2-10keV}=16.7$ (see e.g., \citealt{lusso12,duras20}) and assume that all black holes in our sample accrete at $1\%$ of their Eddington ratio, \revI{these choices are commonly used in the literature, however, we point the reader attention to the fact that they are calibrated on larger mass BHs.} The inferred black hole masses are in the range of $\sim10^4-10^7 \ \rm{M_{\odot}}$. Figure~\ref{fig:mhms} shows the $M_\bullet-M_\star$ relation for the AGN in our sample. The blue line and shaded region indicate the local relation for AGN and its dispersion \citep{reines15}. The red and green lines and shaded regions show the high-redshift relations proposed by \citet{pacucci23} and \citet{li24b}, respectively.

Our sample of AGN in dwarf galaxies is compatible with the low-redshift $M_\bullet-M_\star$ relation, although it shows significant scatter and is biased towards higher luminosities at the low-mass end of the plot (the dotted line indicates the lower limit on the detectable black hole mass). This suggests that local AGN hosted in dwarf galaxies share the same $M_\bullet-M_\star$ relation as more massive $z=0$ galaxies, in agreement with previous results \citep{schutte19,ellis24}.

\section{Comparison with semi-analytical models} \label{sec:mod}

Semi-analytical models (SAM) represent a powerful tool for studying BH seeding mechanisms as well as for exploring growth modes of BHs \citep{somerville08,bower10,ricarte18a,ricarte18b}. In this work, we compare the \revI{properties of the sources in our sample} with mock catalogs of AGN built following the approach described in \citet{chadayammuri23} based on the SAMs developed by \citet{ricarte18b}. 

These SAMs consider two seeding mechanisms and three different mass growth modes, combined to build six mock catalogs.  \revI{In this model, dark matter merger trees are constructed for a variety of halo masses for $0<z<20$.  SMBHs are seeded at $z>15$ with} either rare ``heavy'' seeds, typically resulting from direct collapse (DC) scenarios, distributed following \citet{lodato06} and \citet{volonteri10}, or abundant ``light'' seeds with a distribution motivated by cosmological simulations \citep{hirano15}. This second scenario is motivated by cosmological simulations where BHs are produced by the death of population III (PopIII) stars and the mass distribution of the seeds is represented by a power law spanning from 30 to 100 solar masses \citep{hirano15}. \revI{Most growth is merger-driven: major mergers cause SMBHs to grow to the $M_\bullet-\sigma$ relation, and a few ``steady'' modes are explored to prescribe accretion rates at other times, detailed below.}

\begin{itemize}
    \item Power Law (PL):  When a major merger occurs, central BHs grow at the Eddington rate until they reach the $M_\bullet-\sigma$ relation. Otherwise, they accrete according to a universal Eddington ratio distribution function.
    \item AGN Main Sequence (AGNms):  When a major merger occurs, central BHs grow at the Eddington rate until they reach the $M_\bullet-\sigma$ relation. Otherwise, they accrete at $10^{-3}$ times the star formation rate.
    \item Broad Line Quasar (BLQ):  When a major merger occurs, central BHs grow at Eddington ratios drawn from observed broad line quasars, allowing occasional super-Eddington accretion. Otherwise, the accretion rate is zero.
\end{itemize}

\revI{Previously, \citet{ricarte18b} showed that the PL model produced reasonable AGN luminosity functions for $0 < z < 6$, while the observationally motivated AGNms produced too many faint AGN and too few bright AGN at low redshift.  All models reproduce the $M_\bullet-\sigma$ relation at high mass by construction.  The BLQ model produced even better agreement with AGN luminosity functions \citep{ricarte19}.  However, \citet{chadayammuri23} revealed that this model produced too few faint AGN, due to its lack of a steady mode.  Fundamentally, this is due to the fact that luminosity function constraints do not directly test how given AGN luminosities are assigned to individual galaxies.  As we shall show, our more detailed study probing luminosity as a function of stellar mass illuminates weaknesses in these models.}

\begin{figure}[t!]
\centering
\includegraphics[width = 0.5\textwidth]{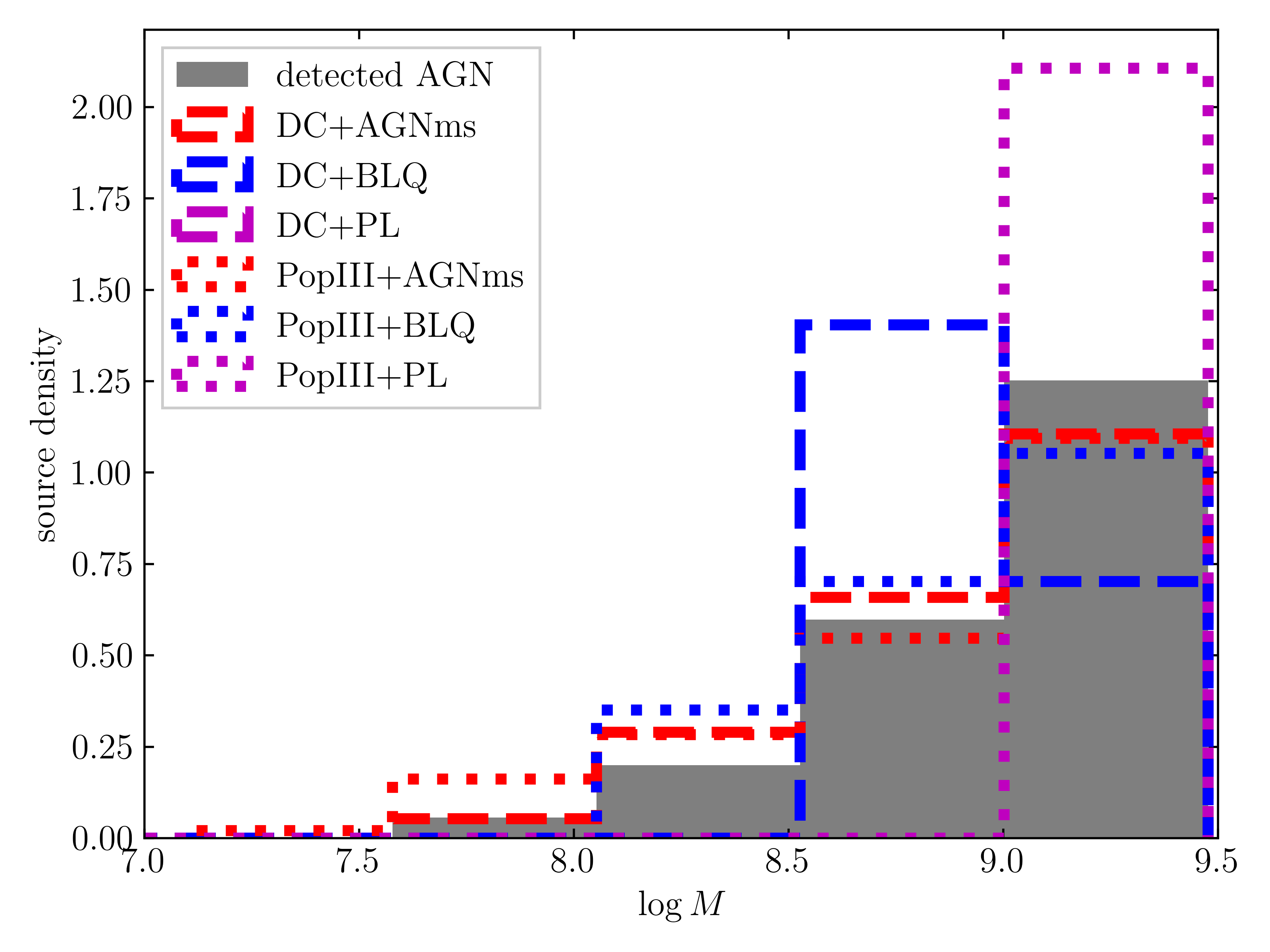}
\caption{Normalized distribution of the AGN host galaxies masses for our sample (in grey) and the six mock catalogs obtained from the described SAM (dashed colored lines).
\label{fig:mass}}
\end{figure}

We can probe growth modes and seeding mechanisms by comparing our observed AGN properties with the mock catalogs built upon the described SAMs. Following \citet{chadayammuri23}, for each SAM, we build the corresponding mock catalog by computing the expected X-ray luminosity based on the mass and mass accretion rate of each BH. We assume an efficiency in converting mass accretion rate into luminosity of $10\%$ and a bolometric-to-X-ray luminosity correction of $10\%$, which are both common and adequate choices (see e.g., \citealt{lusso12,duras20}). \revI{Although these quantities are poorly understood in dwarf galaxies, these are not the leading sources of theoretical uncertainty.} We then filter the obtained mock catalogs using the same luminosity, mass, and distance criteria adopted in the construction of our AGN sample in dwarf galaxies. This approach allows for a direct comparison between the filtered mock catalogs and our observed sample.

\begin{figure}[t!]
\centering
\includegraphics[width = 0.5\textwidth]{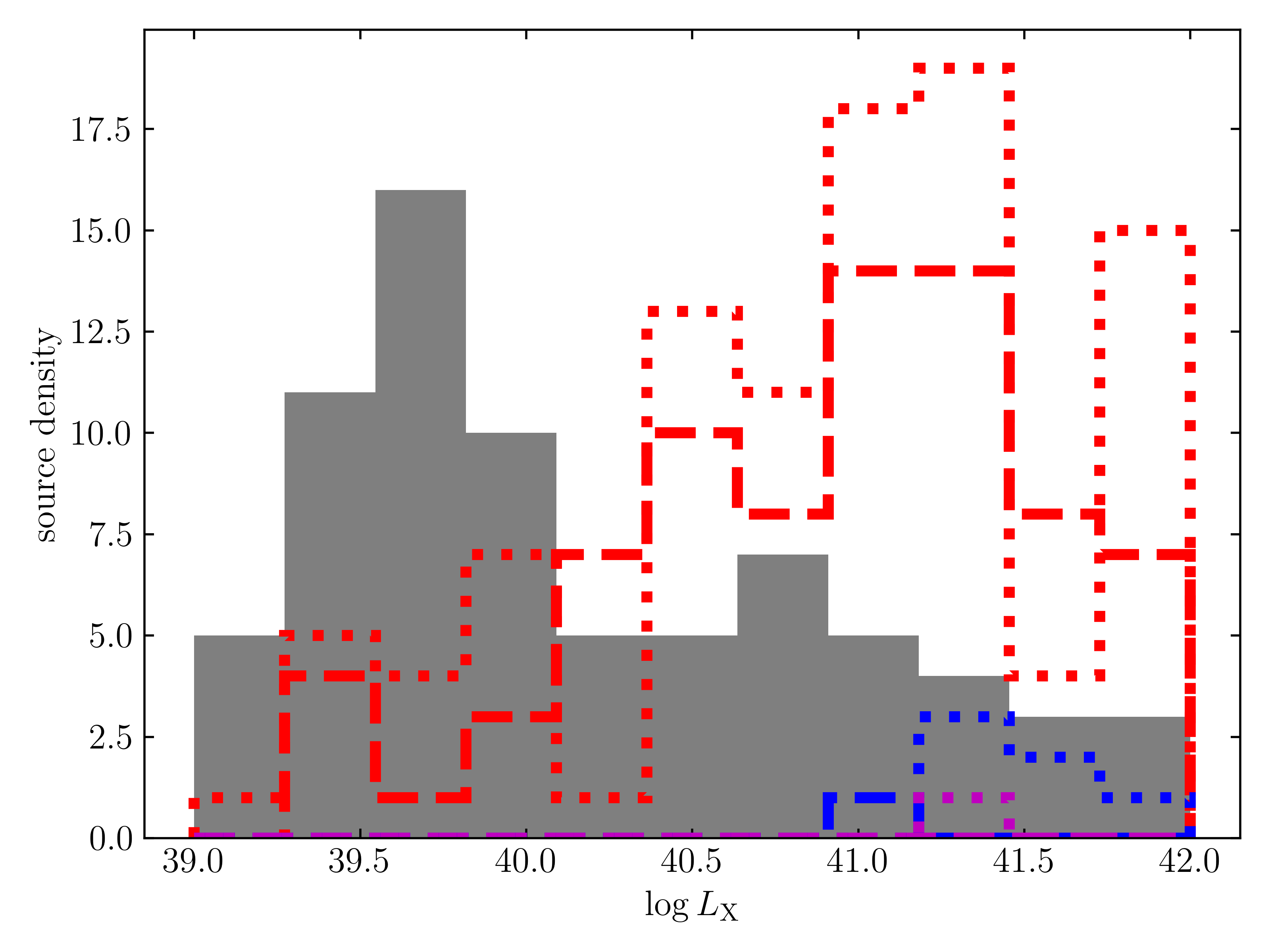}
\caption{Normalized distribution of the AGN X-ray luminosity of our sample (in grey) and the six mock catalogs obtained from the described SAM (dashed colored lines).
\label{fig:lum}}
\end{figure}

Figures \ref{fig:mass} and \ref{fig:lum} present our main results, showing the normalized distributions of host-galaxy masses and luminosities for our sample (in grey) and the six mock catalogs built on different models (colored dashed lines). Three models roughly reproduce the distribution of observed host-galaxy mass: DC+AGNms, PopIII+AGNms, and PopIII+BLQ. \revI{These results suggest that the data prefer models with more abundant AGN in dwarf galaxies. The AGNms model keeps all SMBHs accreting at a moderate rate. While the BLQ model does not have a steady accretion mode, the higher abundance of PopIII seeds keeps the AGN fraction higher than for DC seeds.}

\revI{On the other hand}, the PopIII+BLQ model fails to predict the observed distribution of AGN luminosities, \revI{producing AGN that are too bright.  Again, this points towards the need for a steady mode of AGN activity independent of galaxy mergers to produce fainter AGN at a greater abundance.}  The two best-performing models (DC+AGNms and PopIII+AGNms) \revI{do produce a wider distribution of luminosities, but they} fail to reproduce the observed luminosity distribution. They predict distributions skewed toward higher luminosities, peaking at about $10^{41.5} \ \rm{erg \ s^{-1}}$, while the observed distribution peaks roughly at $10^{39.5} \ \rm{erg \ s^{-1}}$. This suggests that the preferred accretion mode is the one adopted in the AGNms model, but it also highlights the need for more theoretical and modeling efforts to accurately reproduce the observed AGN luminosity distribution. \revI{Recall that the AGNms model universally assigns accretion rates at $10^{-3}$ times the SFR, with a duty cycle of unity. Redistributing this accretion in a distribution in future work may help match these observations.  This discrepancy is missed when calibrating to the luminosity function of more luminous AGN.}

These results offer intriguing insights into the physics of SMBHs and their growth history. Dwarf galaxies typically experience fewer major mergers in their recent assembly compared to their more massive counterparts, making them more sensitive to the ``steady mode'' of accretion. Our results suggests that a good model for this steady mode could involve accretion tied to the star-formation rate of the host galaxy, \revI{or should otherwise keep SMBH active at a moderate rate in dwarf galaxies}.

The results by \citet{chadayammuri23} were obtained using the AMUSE survey \citep{gallo08,miller15b}, a volume-limited survey aimed at identifying AGN in early-type galaxies. \citet{chadayammuri23} concluded that data from the AMUSE survey favor the ``heavy-seed'' scenario but are too shallow to completely rule out the ``light seed'' scenario. This conclusion is further confirmed by our analysis. While we can effectively discriminate between accretion modes, our X-ray-selected data sample is too limited to probe different BH seeding mechanisms. This picture will not change significantly with the release of additional eROSITA scans: although the average detection sensitivity will improve by a factor of $\approx2.5$, the low-luminosity tail of the AGN population will remain inaccessible. In this luminosity regime, LMXBs, HMBXs and ULXs will dominate, and the poor angular resolution of eROSITA does not allow for their unambiguous identification.

\section{Conclusions} \label{sec:conc}

Harnessing the recent release of eROSITA data of the Western part of the sky (which covers 20,000~$\rm{deg^2}$, we compiled a large and uniform sample of X-ray-detected AGN in local dwarf galaxies. After excluding all possible contaminants, our final sample consists of 74 X-ray source-dwarf galaxy pairs. 

We characterized the sources in our sample by investigating their X-ray hardness ratio, their location within the host galaxy, luminosity function, and galaxy-to-BH mass relation. The average X-ray hardness ratio of our sample is compatible with a power-law model with a slope of $\Gamma=1.8$, as expected from a population of AGN. We found that a substantial fraction ($\approx50\%$) of the sources are likely off-nuclear, in line with theoretical predictions. We also constructed a precise X-ray luminosity function for AGN in dwarf galaxies, which we fit with a power-law function. Finally, we compared the BH mass of our sources (inferred from their X-ray luminosity) with their host galaxy stellar mass. We found that the ratio of BH-to-galaxy mass for our sample is in good agreement with the local relation for AGN hosted in more massive galaxies.

To place our results in context, we compared the mass and luminosity distributions of our sample of AGN in dwarf galaxies with predictions from semi-analytical models. Based on this analysis we concluded that the preferred models keep SMBH in dwarf galaxies active at a moderate rate. Among the models tested, this corresponds to one in which the accretion rate is kept proportional to the star-formation rate of the host galaxy.  However, even these models do not reproduce the observed distribution of AGN luminosities in detail, producing too few AGN with $\log L_x \sim 10^{39.5}$ and too many AGN with $\log L_x \sim 10^{41}$, motivating additional modeling efforts to better describe the behavior of AGN in dwarf galaxies. These discrepancies can be missed by modelers calibrating only to luminosity functions of more luminous AGN. Unfortunately, due to the degeneracies between seeding and accretion rates and the shallowness of our sample, we cannot convincingly distinguish between different seeding mechanisms. In addition to improved modeling of AGN in dwarf galaxies, this highlights the necessity of deeper X-ray surveys to convincingly determine the origin of the observed SMBHs.


\section*{Acknowledgments}
\'A.B. acknowledges support from the Smithsonian Institution and the Chandra Project through NASA contract NAS8-03060. This work was supported in part by the Black Hole Initiative at Harvard University, made possible through the support of grants from the Gordon and Betty Moore Foundation and the John Templeton Foundation. The opinions expressed in this publication are those of the author(s) and do not necessarily reflect the views of the Moore or Templeton Foundations.

%







\bibliography{biblio}{}
\bibliographystyle{aasjournal}

\appendix
\section{Properties of the selected sources}
We report in Table~\ref{tab:prop} the properties of the 74 AGN in dwarf galaxies we selected.

\begin{table*}
\begin{center}
\rotatebox{90}{
\begin{minipage}{\textheight}
\caption{Properties of the 74 AGN selected.}\label{tab:prop}
\setlength{\leftskip}{-40pt}
\footnotesize
\begin{tabular}{lccccccccc}
\hline\hline
 ID & R.A. & DEC & R.A.$_\textup{X}$ & DEC$_\textup{X}$ & $\sigma_p$ & $D$ [Mpc] &$\log(M_\star/{\rm M_\odot})$ & $L_\textup{X}~[10^{40}~{\rm erg/s}]$ & Nuclear\\
\\
 \hline
PGC135069 & 52.2929 & -22.0213 & 52.2929 & -22.0216 & 0.97 & 24.5 & 9.41 & $ 6.55 \pm 0.47 $& 0 \\
WINGSJ043139.17-612330.1 & 67.9132 & -61.3916 & 67.9166 & -61.3901 & 1.87 & 41.03 & 8.22 & $ 11.37 \pm 0.67 $& 0 \\
PGC200277 & 179.382 & 32.3408 & 179.3816 & 32.3399 & 3.98 & 48.36 & 9.02 & $ 3.1 \pm 1.09 $& 1 \\
6dFJ1551458-091534 & 237.9408 & -9.2594 & 237.9406 & -9.2597 & 2.01 & 75.19 & 8.87 & $ 24.03 \pm 3.99 $& 1 \\
ESO411-012 & 11.7126 & -31.541 & 11.7127 & -31.5383 & 4.89 & 19.71 & 7.66 & $ 0.22 \pm 0.09 $& 0 \\
PGC2822735 & 25.2132 & -32.7364 & 25.2124 & -32.7355 & 3.96 & 166.27 & 9.45 & $ 64.99 \pm 12.69 $& 0 \\
6dFJ0142532-432355 & 25.7217 & -43.3986 & 25.7214 & -43.3983 & 1.83 & 19.22 & 7.78 & $ 1.72 \pm 0.21 $& 1 \\
NGC1034 & 39.5584 & -15.8091 & 39.5571 & -15.8082 & 5.05 & 19.28 & 9.15 & $ 0.27 \pm 0.09 $& 0 \\
6dFJ0245228-240617 & 41.3451 & -24.1046 & 41.3445 & -24.1048 & 1.39 & 101.61 & 9.24 & $ 72.89 \pm 7.01 $& 0 \\
ESO479-025 & 40.533 & -24.1292 & 40.533 & -24.1275 & 5.09 & 16.9 & 9.31 & $ 0.19 \pm 0.07 $& 0 \\
PGC3213080 & 42.5881 & -29.5708 & 42.5866 & -29.571 & 6.68 & 76.4 & 8.76 & $ 2.57 \pm 1.06 $& 1 \\
PGC012154 & 49.0887 & -25.8548 & 49.0894 & -25.8546 & 5.3 & 19.66 & 9.37 & $ 0.25 \pm 0.08 $& 1 \\
PGC693568 & 47.0262 & -32.3275 & 47.0256 & -32.327 & 3.28 & 65.84 & 8.75 & $ 5.14 \pm 1.12 $& 1 \\
PGC540963 & 50.5536 & -44.0946 & 50.5543 & -44.0946 & 5.37 & 122.51 & 9.28 & $ 3.66 \pm 1.71 $& 1 \\
PGC134076 & 53.7236 & -24.7909 & 53.7241 & -24.7917 & 4.62 & 21.76 & 8.45 & $ 0.19 \pm 0.08 $& 1 \\
6dFJ0342278-260243 & 55.6158 & -26.0452 & 55.6171 & -26.0459 & 11.14 & 21.4 & 8.23 & $ 0.18 \pm 0.08 $& 1 \\
PGC3315038 & 63.0987 & -54.4213 & 63.0968 & -54.4219 & 6.25 & 173.09 & 9.47 & $ 4.92 \pm 2.21 $& 0 \\
6dFJ0421194-313805 & 65.3308 & -31.6346 & 65.3303 & -31.6341 & 1.28 & 32.95 & 8.67 & $ 8.19 \pm 0.78 $& 0 \\
PGC126927 & 66.4561 & -61.1529 & 66.4528 & -61.1535 & 5.46 & 68.36 & 8.88 & $ 0.75 \pm 0.31 $& 0 \\
ESO118-034 & 70.0728 & -58.7443 & 70.0717 & -58.7467 & 3.37 & 13.65 & 9.27 & $ 0.21 \pm 0.03 $& 0 \\
PGC928553 & 73.1952 & -14.1312 & 73.1948 & -14.1297 & 3.67 & 39.15 & 9.37 & $ 0.95 \pm 0.35 $& 0 \\
PGC016675 & 76.1302 & -16.5847 & 76.1323 & -16.5854 & 5.16 & 43.61 & 9.37 & $ 0.59 \pm 0.31 $& 0 \\
PGC016986 & 78.8377 & -26.468 & 78.8379 & -26.4709 & 3.8 & 50.47 & 8.71 & $ 1.38 \pm 0.52 $& 0 \\
UGC03282 & 79.4064 & 6.7994 & 79.4075 & 6.8012 & 5.59 & 80.21 & 9.42 & $ 3.47 \pm 1.64 $& 0 \\
6dFJ0532495-322925 & 83.2062 & -32.4904 & 83.2047 & -32.4939 & 4.64 & 148.13 & 9.47 & $ 10.68 \pm 4.07 $& 0 \\
PGC917425 & 84.0581 & -14.977 & 84.0584 & -14.9759 & 4.91 & 26.12 & 8.82 & $ 0.62 \pm 0.2 $& 1 \\
2MASXJ05403594-5436417 & 85.1497 & -54.6116 & 85.1542 & -54.6139 & 3.98 & 32.35 & 9.43 & $ 0.38 \pm 0.11 $& 0 \\
PGC148318 & 90.956 & -33.1912 & 90.9581 & -33.1915 & 5.65 & 36.89 & 9.17 & $ 0.54 \pm 0.36 $& 0 \\
PGC075555 & 93.0592 & -38.7731 & 93.0568 & -38.7732 & 4.78 & 21.07 & 8.43 & $ 0.2 \pm 0.09 $& 0 \\
PGC483594 & 92.438 & -48.7986 & 92.4384 & -48.8023 & 4.04 & 122.17 & 9.27 & $ 10.04 \pm 2.95 $& 0 \\
PGC019486 & 100.792 & -76.5594 & 100.7784 & -76.5586 & 5.15 & 50.73 & 9.13 & $ 0.43 \pm 0.16 $& 0 \\
PGC088580 & 97.7514 & -56.6482 & 97.7462 & -56.6468 & 10.4 & 142.62 & 9.41 & $ 25.28 \pm 6.02 $& 0 \\
PGC747791 & 102.7671 & -27.9423 & 102.7673 & -27.9426 & 4.24 & 31.13 & 9.1 & $ 0.93 \pm 0.32 $& 1 \\
PGC2054090 & 120.4398 & 34.7579 & 120.4402 & 34.7612 & 6.77 & 65.89 & 9.31 & $ 2.13 \pm 1.27 $& 0 \\
PGC023484 & 125.6199 & -1.1028 & 125.6198 & -1.1023 & 1.9 & 62.38 & 9.16 & $ 34.09 \pm 4.47 $& 1 \\
PGC3093997 & 130.9451 & -17.6794 & 130.944 & -17.6782 & 6.28 & 22.81 & 9.34 & $ 0.43 \pm 0.2 $& 1 \\
PGC024469 & 130.7008 & 14.2652 & 130.6996 & 14.2665 & 7.11 & 29.99 & 9.45 & $ 0.44 \pm 0.26 $& 1 \\
PGC153352 & 134.7511 & -18.3842 & 134.7513 & -18.3851 & 4.46 & 26.86 & 9.25 & $ 0.6 \pm 0.23 $& 1 \\
NGC2777 & 137.6744 & 7.2067 & 137.6738 & 7.2045 & 4.51 & 25.7 & 9.33 & $ 0.74 \pm 0.27 $& 0 \\
IC0549 & 145.1801 & 3.9592 & 145.1792 & 3.9598 & 6.69 & 22.7 & 8.93 & $ 0.45 \pm 0.2 $& 1 \\
PGC2077238 & 145.8306 & 36.2369 & 145.8302 & 36.2392 & 5.73 & 87.52 & 9.46 & $ 4.91 \pm 2.53 $& 0 \\
AGC193818 & 147.9617 & 13.9109 & 147.961 & 13.9144 & 14.23 & 98.06 & 9.39 & $ 31.27 \pm 9.12 $& 1 \\
NGC3125 & 151.6388 & -29.9353 & 151.6411 & -29.9371 & 3.14 & 14.98 & 9.37 & $ 0.49 \pm 0.12 $& 0 \\
\hline\hline\\
\end{tabular}
\normalsize
\end{minipage}
}
\end{center}
\end{table*}

\begin{table*}
\begin{center}
\rotatebox{90}{
\begin{minipage}{\textheight}
\caption{Continued.}\label{tab:prop}
\setlength{\leftskip}{-40pt}
\footnotesize
\begin{tabular}{lccccccccc}
\hline\hline
 ID & R.A. & DEC & R.A.$_\textup{X}$ & DEC$_\textup{X}$ & $\sigma_p$ & $D$ [Mpc] &$\log(M_\star/{\rm M_\odot})$ & $L_\textup{X}~[10^{40}~{\rm erg/s}]$ & Nuclear\\
\\
 \hline
6dFJ1024202-201458 & 156.0839 & -20.2494 & 156.0842 & -20.2494 & 3.34 & 69.73 & 8.96 & $ 15.63 \pm 3.48 $& 1 \\
IC2604 & 162.3545 & 32.7728 & 162.354 & 32.7749 & 5.56 & 25.1 & 8.97 & $ 0.44 \pm 0.2 $& 0 \\
UGC05989 & 163.1327 & 19.7923 & 163.1322 & 19.7929 & 3.89 & 16.07 & 8.78 & $ 0.51 \pm 0.14 $& 1 \\
PGC096512 & 173.696 & -59.2293 & 173.6943 & -59.2298 & 5.91 & 18.51 & 8.48 & $ 0.13 \pm 0.05 $& 0 \\
NGC4016 & 179.6209 & 27.5288 & 179.6184 & 27.53 & 6.43 & 50.62 & 9.46 & $ 1.6 \pm 0.86 $& 0 \\
ESO321-018 & 183.9762 & -38.0934 & 183.9723 & -38.0909 & 4.25 & 41.79 & 9.14 & $ 0.93 \pm 0.37 $& 0 \\
ESO267-032 & 183.6912 & -43.5649 & 183.6885 & -43.565 & 5.59 & 26.01 & 9.22 & $ 0.22 \pm 0.11 $& 0 \\
2MASXJ12260880+2826008 & 186.5367 & 28.4336 & 186.5365 & 28.4348 & 8.18 & 64.95 & 9.12 & $ 2.16 \pm 1.15 $& 1 \\
PGC039730 & 184.9716 & 1.7734 & 184.974 & 1.7732 & 5.96 & 29.37 & 8.72 & $ 0.39 \pm 0.21 $& 0 \\
PGC039904 & 185.2971 & 17.6386 & 185.2966 & 17.6374 & 4.52 & 16.98 & 8.69 & $ 0.18 \pm 0.08 $& 0 \\
UGC07366 & 184.8695 & 17.2304 & 184.8706 & 17.2303 & 6.63 & 17.52 & 8.92 & $ 0.15 \pm 0.08 $& 1 \\
PGC560327 & 191.9905 & -42.6486 & 191.9906 & -42.6497 & 4.91 & 26.5 & 9.13 & $ 0.24 \pm 0.11 $& 1 \\
ESO324-008 & 200.2248 & -39.2788 & 200.2239 & -39.2774 & 4.72 & 27.12 & 8.99 & $ 0.66 \pm 0.19 $& 0 \\
HIZOAJ1343-65 & 205.8547 & -65.2646 & 205.8548 & -65.2651 & 5.37 & 43.61 & 8.94 & $ 0.88 \pm 0.37 $& 1 \\
LAMOSTJ134630.99+070428.6 & 206.6291 & 7.0746 & 206.6287 & 7.0747 & 2.41 & 110.43 & 9.17 & $ 33.71 \pm 5.98 $& 1 \\
ESO221-009 & 207.686 & -48.9469 & 207.688 & -48.9474 & 4.71 & 41.13 & 8.92 & $ 1.2 \pm 0.4 $& 0 \\
NGC5398 & 210.34 & -33.0637 & 210.3406 & -33.062 & 2.83 & 11.39 & 9.21 & $ 0.23 \pm 0.05 $& 0 \\
ESO271-018 & 212.5152 & -46.2224 & 212.5127 & -46.2235 & 5.01 & 35.29 & 8.15 & $ 0.52 \pm 0.24 $& 0 \\
PGC679199 & 212.6113 & -33.2269 & 212.6111 & -33.2282 & 5.01 & 50.8 & 9.4 & $ 0.97 \pm 0.47 $& 1 \\
PGC538542 & 218.2205 & -44.3216 & 218.2217 & -44.3212 & 4.41 & 15.51 & 8.65 & $ 4.72 \pm 0.32 $& 1 \\
ESO580-005 & 219.5665 & -22.3245 & 219.5662 & -22.3238 & 5.92 & 34.14 & 9.06 & $ 0.59 \pm 0.25 $& 1 \\
PGC812038 & 221.4301 & -22.5793 & 221.4315 & -22.5811 & 4.67 & 47.16 & 9.46 & $ 13.42 \pm 1.87 $& 0 \\
PGC053603 & 225.1221 & -26.4499 & 225.1218 & -26.4511 & 2.8 & 74.12 & 9.46 & $ 7.49 \pm 1.47 $& 0 \\
6dFJ1458243-373012 & 224.6013 & -37.5031 & 224.6022 & -37.5028 & 1.74 & 104.8 & 9.33 & $ 64.95 \pm 7.9 $& 0 \\
PGC329372 & 309.7387 & -63.771 & 309.74 & -63.7704 & 5.39 & 21.35 & 8.44 & $ 0.29 \pm 0.13 $& 1 \\
2MASXJ21390659-4353001 & 324.7777 & -43.8832 & 324.7775 & -43.8832 & 2.65 & 80.99 & 9.47 & $ 22.01 \pm 4.58 $& 1 \\
PGC3321235 & 332.5453 & -56.0747 & 332.5466 & -56.075 & 3.28 & 55.31 & 8.61 & $ 5.99 \pm 1.57 $& 0 \\
PGC3320749 & 330.1861 & -56.6874 & 330.1838 & -56.6863 & 4.92 & 63.14 & 8.81 & $ 2.61 \pm 1.16 $& 0 \\
NGC7657 & 351.698 & -57.8058 & 351.6937 & -57.8063 & 5.7 & 40.41 & 9.37 & $ 0.74 \pm 0.4 $& 0 \\
ESO240-012 & 354.7067 & -51.8597 & 354.7074 & -51.8597 & 5.15 & 23.69 & 9.41 & $ 0.41 \pm 0.16 $& 1 \\
ESO241-006 & 359.0627 & -43.4275 & 359.0647 & -43.4243 & 4.78 & 18.94 & 8.92 & $ 0.29 \pm 0.12 $& 0 \\
\hline\hline\\
\end{tabular}
\normalsize
\textcolor{white}{Identifier, R.A., DEC,} Identifier, R.A., DEC, distance, and galaxy stellar mass were taken from the HECATE catalog. R.A.$_\textup{X}$ and DEC$_\textup{X}$ are the coordinates of the X-ray source associated with the galaxy. $\sigma_\textup{p}$ is the positional error of the X-ray source as reported by the eRASS1 catalog (\texttt{POS$\_$ERR}). The X-ray luminosity is computed assuming the distance reported upon the eRASS1 flux in the $0.2-2.3$~keV band. The {\it Nuclear} flag indicates whether the X-ray source location is compatible with its dwarf galaxy centroid.
\end{minipage}
}
\end{center}
\end{table*}

\section{X-ray archival data}\label{app:xray}

\subsection{Know ULXs}

\subsubsection{ESO~338-004}
This galaxy is well known in the literature as the host of three ULXs, one of which might be an intermediate-mass black hole (IMBH) accretor \citep{oskinova19}.

\subsubsection{SDSS~J121855.02+142445.5}
Another source is hosted in NGC~4254 and it is likewise thought to be powered by accretion onto an IMBH \citep{mezcua13}.

\subsubsection{NGC~4038}
One X-ray source is coincident with the nuclear region of the Antennae galaxy NGC~4038. This region has been imaged with both \xmm\ and \cxo\ and its emission is attributed to the presence of several ULXs \citep{fabbiano01}.

\subsubsection{PGC~014118}
This galaxy location has been visited with \xmm\ in February 2007. The X-ray emission appears to have originated from, at least, three different sources in the \xmm\ image, spatially coincident with the location of a nuclear star cluster (NSC) \citep{fahrion22}. This suggests that the X-ray emission of this galaxy is caused by the blending of multiple ULXs, although the presence of an AGN cannot be ruled out.

\subsection{Followed-up candidate AGN}

\subsubsection{NGC~3125}
This source is a bright infrared LINER with a compact source of hard X-rays in its nucleus which is interpreted as the result of AGN activity \citep{dudik05}.

\subsubsection{IC~549}
This source's location falls in the eFEDS footprint \citep{brunner22} and was already indicated as an AGN candidate \citep{latimer21}. A follow-up observation by \cxo\ in 2022 further confirmed the AGN nature of this source.

\subsubsection{PGC~039904}
This source has been observed with \cxo\ and, given that its location lies outside the nucleus of its host galaxy, it is considered a ULX \citep{thygensen23}, although \citet{bellovary21} predicts that a fraction as large as 50\% of AGN hosted in dwarf galaxies should be removed from its host nuclear region.

\subsection{Bona fide AGN}

\subsubsection{PGC~053603}
This galaxy was observed with \xmm\ in February 2006 for $\approx10$ ks. Its X-ray spectrum is well-modelled by an absorbed power law with photon index $\Gamma=2.3\pm0.2$ and intrinsic absorption $N_\textup{H}=(1.5\pm0.5)\times10^{21}$ cm$^{-2}$. The unabsorbed luminosity of the source in the $2-10$ keV band amounts to $(5.4\pm0.9)\times10^{40} \ \rm{erg \ s^{-1}}$. 

\subsubsection{PGC~538542}
This galaxy was observed three times, in 2003, 2009 and 2014. We focus our analysis on this latest observation which is the longest and cleanest of the three, amounting to more than 90 ks. The X-ray spectrum of the nuclear region of this galaxy is well-modelled by a multi-component spectrum composed of a thermal plasma a power law and an emission Gaussian line at 5.9 keV. The thermal plasma temperature is $kT=1.17\pm0.04$ keV, and the power law slope is $\Gamma=2.20\pm0.08$. The unabsorbed luminosity of the sole power-law 
component in the $2-10$ keV band amount to $(3.7\pm0.2)\times10^{39} \ \rm{erg \ s^{-1}}$.

\subsubsection{IC~2604}
This galaxy has been observed with both \xmm\ and \cxo\ and, although in both instruments its location falls more than $10'$ from the observation pointing, the source is detected in both cases and the source location is roughly coincident with the nuclear region of its host galaxy. Its X-ray spectrum is well modelled by a power law with photon index $\Gamma=1.9\pm0.1$ and the unabsorbed luminosity in the 2--10 keV band amounts to $L_\textup{X}=(5\pm2)\times10^{39} \ \rm{erg \ s^{-1}}$.

\subsubsection{PGC~134076}
This source has been serendipitously observed with \cxo. Its X-ray spectrum can be reproduced by a power law with photon index $\Gamma=1.5\pm0.2$ and its X-ray luminosity in the 2--10 keV band is $L_\textup{X}=(4\pm1)\times10^{39} \ \rm{erg \ s^{-1}}$.

\subsubsection{NGC~4016}
The source location falls on the border of \xmm\ detectors in a short and heavily affected by flaring observation in November 2020. Although the source is detected the handful of counts obtained does not allow for building a spectrum.

\section{Mock catalogs properties}
\begin{figure*}[t!]
\centering
\includegraphics[width = 0.9\textwidth]{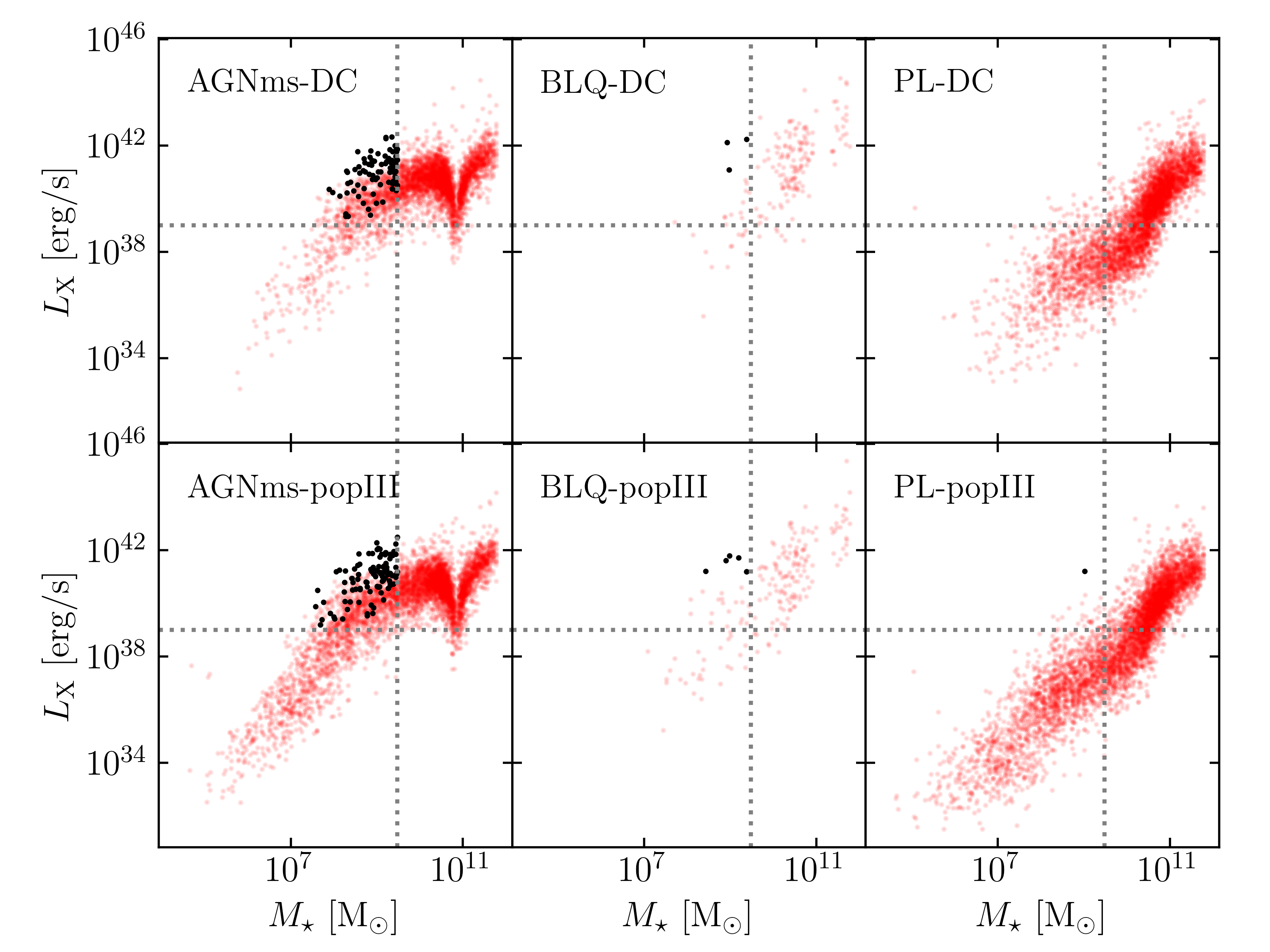}
\caption{X-ray luminosity as a function of the stellar mass for each mock catalog we compared to our AGN sample. Shaded red dots indicate all the sources in our catalogs, and black dots are the ones that pass our selection criteria. The grey dotted lines show the threshold of our selection: no source with a luminosity below $10^{39} \ \rm{erg \ s^{-1}}$ or with a mass larger than $3\times10^9~{\rm M}_\odot$ was considered.
\label{fig:xlum}}
\end{figure*}

The mock catalogs we built are based on the SAMs described in \citet{ricarte18b}, following the same approach of \citet{chadayammuri23}. Figure \ref{fig:xlum} shows, for each SAM, the X-ray luminosity as a function of the stellar mass, and the cuts we imposed building our sample. 

\end{document}